\documentclass[fleqn,usenatbib]{mnras}
\usepackage{newtxtext,newtxmath}
\usepackage{mathptmx}
\usepackage{txfonts}

\usepackage[T1]{fontenc}

\DeclareRobustCommand{\VAN}[3]{#2}
\let\VANthebibliography\thebibliography
\def\thebibliography{\DeclareRobustCommand{\VAN}[3]{##3}\VANthebibliography}

\usepackage{graphicx}	
\usepackage{amsmath}	
\usepackage{amssymb}	
\usepackage{hyperref}
\usepackage{footnote}

\title[Homologous Surges]{Diagnostic of Homologous Solar Surge Plasma\\ as observed by {\it IRIS} and {\it SDO}}

\author[Kayshap et al.]{
Pradeep Kayshap\thanks{e-mail:virat.com@gmail.com},
Rajdeep Singh Payal,
Sharad C. Tripathi, 
and Harihara Padhy 
\\
VIT Bhopal University, Bhopal - Indore Highway, Kothri Kalan, Sehore - 466114, M.P., India\\
}

\date{Accepted XXX. Received YYY; in original form ZZZ}

\pubyear{2021}

\begin{document}
\label{firstpage}
\pagerange{\pageref{firstpage}--\pageref{lastpage}}
\maketitle

\begin{abstract}

Surges have regularly been observed mostly in H$_{\alpha}$ 6563~{\AA} and Ca~{\sc ii} 8542~{\AA}. However, surge$'$s response to other prominent lines 
of the interface-region (Mg~{\sc ii} k 2796.35~{\AA} $\&$ h 2803.52~{\AA}, O~{\sc iv} 1401.15~{\AA}, Si~{\sc iv} 1402.77~{\AA}) is not well studied. 
Here, the evolution and kinematics of six homologous surges are analysed using IRIS and AIA observations. These surges 
were observed on 7$^{th}$ July 2014, located very close to the limb. DEM analysis is performed on these surges 
where the co-existence of cool (log T/K = 6.35) and relatively hot (log T/K = 6.95) components have been found 
at the base. This demonstrates that the bases of surges undergo substantial heating. During the emission of these surges in the above mentioned 
interface-region lines, being reported here for the first time, two peaks have been observed in the initial phase of emission,
where one peak is found to be constant while other one as varying, i.e., non-constant (observed red to
blueshifts across the surge evolution) in nature. This suggests the rotational motion of surge plasma. The heated base and 
rotating plasma suggests the occurrence of magnetic reconnection, most likely, as the trigger for homologous surges. During the emission 
of these surges, it is found that despite being optically thick (i.e., R$_{kh}$ < 2.0), central reversal was not observed for Mg~{\sc ii} k $\&$ h 
lines. Further, R$_{kh}$ increases with surge emission in time and it is found to have positive correlation with Doppler Velocity 
while negative with Gaussian width.

\end{abstract}




\section{Introduction}
Jet-like features (spicules, chromospheric anemone jets, macrospicules, surges, X-ray/Ultraviolet (UV)/Extreme UV 
(EUV) jets, etc.) are ubiquitous in the solar atmosphere. 
Study of these jet-like structures is one of the important areas in solar physics research. Research
in this field is continuously increasing which adds on to our knowledge about various aspects of these jet-like structures, 
e.g., formation, evolution, plasma properties (e.g., \citealt[and references cited therein]{Bohlin1975, 
Canfield1996,Yokoyama1996,Innes1997,Shibata2007,Nisizuka2008,Abhi2011,Kayshap2013a,Kayshap2013b,Jelinek2015,
Cheung2015b,Kayshap2018a}).

Solar surges are cool plasma structures that typically emit in the H$_{\alpha}$ and other Chromospheric/Coronal 
lines (\citealt{Sterling2000}). The Flaring Region, and sites of transient and dynamical activities are the favorable 
sites for surge's origin. Therefore, the knowledge about dynamics of magnetic field is a key factor to understand the 
origin of surges. Various processes (e.g., emerging flux-region, moving magnetic features, etc.) involve the dissipation of magnetic energy. 
This, leads to the formation of surges (e.g., \citealt{Sch1984,Chae1999,Brooks2007,Uddin2012,
Kayshap2013a}). Surges are most likely triggered by magnetic reconnection (a process which converts magnetic energy 
into thermal energy) and exhibit different episodes of heating and cooling (\citealt{Brooks2007}).

A significant amount of the research has been done, so far, to understand the various properties of solar surges, e.g., 
projected height, line-of-sight (LOS) apparent velocity range, life-time, plasma up-flow/downfall projected speeds, 
temperature, and plasma density etc.,(e.g., \citealt{Roy1973,Sch1984,Chae1999, Sterling2000,Gugli2010,Uddin2012,
Kayshap2013a,Raouafi2016}). Such parameters in conjunction with energy release process(es) are important to 
numerically simulate jet-like structures. And, the numerical simulations help to reveal the underlying physics of the jet-like 
structures (e.g., \citealt{Yokoyama1996,Nisizuka2008,Pariat2009,Pariat2010,Kayshap2013a,Kayshap2013b,
Cheung2015b,Raouafi2016}). Usually surges are recurrent in nature, i.e., often several surges 
originate from the same location (e.g., \citealt{Sch1995,Gaiz1996,Jiang2007,Uddin2012}). The reason of such recurrency of surges is, most likely, the 
multiple reconnections at the origin site having high-level complexity in magnetic field configuration in the vicinity of origin site.

Surges are found to be associated with various other features also, e.g., emerging flux-regions (\citealt{Kurokawa2007}), 
light bridge (\citealt{Asai2001,Robustini2016}), and EUV or X-ray ejections (\citealt{Canfield1996,Chen2008,Maria2009,
Zhang2014}). The association of surges with EUV or X-ray emission (i.e., hot component) is one 
of it's important features and multiple efforts have been made, so far, to understand their inter-relationship (e.g., \citealt{Sch1988, 
Shibata1992,Yokoyama1996,Shimojo1996,Liu2004}). In addition, the presence of helical/rotating motion in surges has
also been reported (e.g., \citealt{Gu1994,Canfield1996,Jibben2004}). Furthermore, numerical experiments reveal that 
helical/rotating motion is an indirect signature of magnetic reconnection, i.e., helical magnetic field lines are 
produced through magnetic reconnection between the twisted and pre-existing magnetic field. The plasma flows 
in such helical magnetic field lines, this resembles as rotating plasma (e.g., \citealt{Fang2014, Pariat2016}).

Recently, with the help of Interface Region Imaging Spectrograph (IRIS; \citealt{DePon2014}), surge emission in 
Si~{\sc iv} line was reported, for the first time, by \citealt{Nob2017}. Traditionally, surges are being studied in the 
chromospheric H$_{\alpha}$ and Ca~{\sc ii} H $\&$ K lines. 
It is important to note that IRIS provides observations in several other prominent lines e.g., Mg~{\sc ii}, Si~{\sc iv},
O~{\sc iv}, etc.) of the interface-region which may lead to the crucial informations about surges and in-situ physical 
conditions, as well. Estimation of the ratio of Mg~{\sc ii} k over h (i.e., R$_{kh}$) became popular after the launch 
of IRIS in 2013, e.g., estimation of R$_{kh}$ in the flaring region (\citealt{Kerr2015}), filament region (\citealt{Hara2014}), 
and coronal mass ejections (\citealt{Zhang2019}). R$_{kh}$ is an important parameter to understand the 
formation of spectral profiles of Mg~{\sc ii} lines, and associated in-situ physical conditions (\citealt{Rubio2017}).

This work aims to understand the kinematics, thermal structures, and response of the chromosphere and TR during six homologous 
surges. We have utilized IRIS and Atmospheric Imaging Assembly (AIA) (\citealt{Lemen2012}) observations to explore various aspects 
of these homologous surges. The paper is organized as follows: Section~\ref{sect:obs_data} describes the observations and data 
analysis techniques, Section~\ref{sect:results} describes the results. And, finally, the discussion and conclusions are  
presented in Section~\ref{sect:discussion}. 
 
\section{Observations and Data Analysis}
\label{sect:obs_data}
This work utilizes the imaging and spectroscopic observations. IRIS provides high-resolution spectroscopic $\&$ imaging 
observations over a wide range of the solar atmosphere (\citealt{DePon2014}). The Atomospheric Imaging Assembly (AIA) 
onboard Solar Dynamics Observatory (SDO) provides high-resolution full-disk observations of Sun using various filters that 
capture the solar emission in different temperature ranges (e.g., \citealt{Lemen2012}). \\
The spatial resolution of IRIS slit jaw images (SJI) is 0.33 arcsec. The baseline cadence of 
IRIS/SJI is 5s. In the case of IRIS spectroscopic observations, the slit width is 0.33 arcsec while spatial resolution 
along the slit (i.e., y-direction) is 0.16 arcsec (\citealt{DePon2014}). IRIS spectral resolution is 12.8 m{\AA} for 
far-ultraviolet and 25.6 m{\AA} for near-ultraviolet emissions. The baseline cadence for IRIS spectral observations is
1s (https://iris.gsfc.nasa.gov/spacecraft$\_$inst.html). However, the spatial resolution of AIA is 0.6 arcsec (\citealt{Lemen2012}), 
which is relatively lower as compared to IRIS/SJI spatial resolution. AIA samples the emission in various wavelengths at cadence 
of 10 to 12 seconds as mentioned by \cite{Lemen2012}. By employing these high-end observational capabilities of IRIS and AIA, we have taken
the respective observations of homologous surge emissions with the cadence of 32 seconds in case of IRIS/SJI, and the 16 s cadence
in case of spectral observations. AIA observations with the cadence of 12 s have been employed for the present analysis.
%

Both the space observatories have observed six homologus surges, taken for present study, on 7$^{th}$ July 2014 during 11:19:43~UT to 15:27:54~UT. 
Level 2 data files provided by the IRIS and AIA are standard scientific objects. These data files are ready to use for the scientific analysis. 
For the present work, we have utilized four IRIS spectral lines, namely, Mg~{\sc ii} k 2796.35~{\AA}, Mg~{\sc ii} h 2803.52~{\AA},
Si~{\sc iv} 1402.77~{\AA} and O~{\sc iv} 1401.15~{\AA}. Mg~{\sc ii} k 2796.35~{\AA} emission corresponds to chromosphere while 
Si~{\sc iv} 1402.77~{\AA} O~{\sc iv} 1401.15~{\AA} corresponds to emission from TR.

Single Gaussian fit apparoch has been implemented to deduce the various spectroscopic parameters, e.g., Intensity, Doppler Velocity, and 
Full Width Half Maximum (FWHM). Generally, Mg~{\sc ii} k 2796.35~{\AA} emission is found in optically thick atmosphere and got observed having complex 
emission profiles depending on the in-situ plasma conditions (i.e., two or more than two peaks: 
\citealt{Peter2014,Kayshap2018b}). However, this particular observation of AR 12114 lies outside the limb, and majority of Mg~{\sc ii} 
k 2796.35~{\AA} emissions exhibit single peak. Thus, single Gaussian fit approch is more suitable to decuce the various 
spectrosopic parameters. The deduced intensity, Doppler velocity and FWHM maps of Mg~{\sc ii} and Si~{\sc iv} lines are shown 
in the appendix~\ref{append:maps} (see, figure~\ref{fig:mg_map_append} and~\ref{fig:siv_map_append}). Further, 
to investigate the thermal nature of surge plasma, we have performed Emission Measure (EM) and Differntial Emission Measure
(DEM) analysis using the software provided by \citealt{Cheung2015}. 

\section{Results}
\label{sect:results}
\begin{figure*}
    \mbox{
    \includegraphics[trim = 2.5cm 1.8cm 2.5cm 2.0cm,scale=0.80]{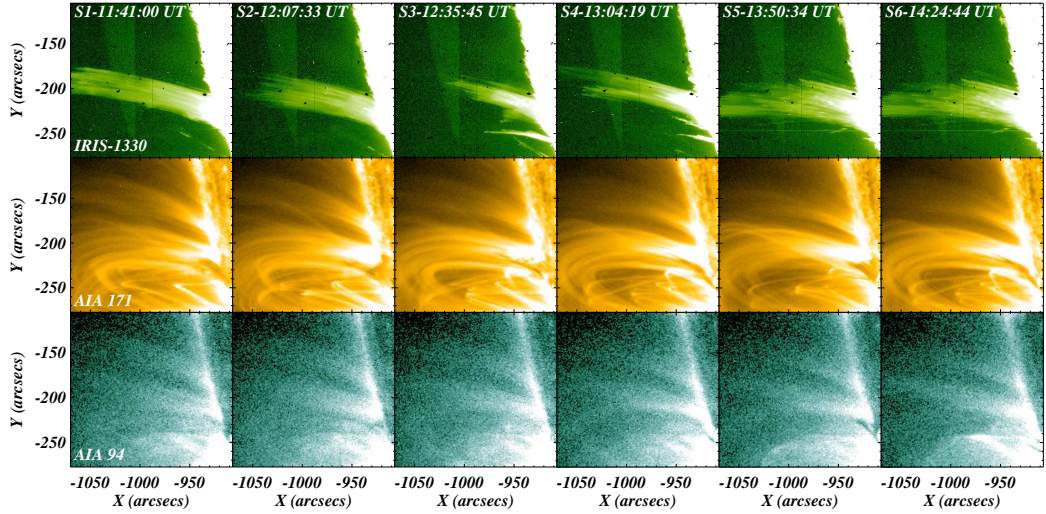}
    }
    \caption{Homologous six surges during the peak time in IRIS 1330~{\AA} (top-row), AIA 171~{\AA} (middle-row), 
	and AIA 94~{\AA} (bottom-row). The surges are clearly visible in IRIS/1330, moderately visible in AIA~171{\AA}, 
	and rarely visible in AIA~94~{\AA}.}
    \label{fig:ref_image}
\end{figure*}

The AR 12114, located at the East limb, produced six homologus surges on 7$^{th}$ July 2014 within the observation time of $\sim$ 4 hours 
(i.e., from 11:19:43~UT to 15:27:54~UT). Fortunately, this dynamic event was captured by both the AIA $\&$ IRIS instruments, simultenously. 
This allowed us to perform the detailed diagnostics of this dynamic event. Fig.~\ref{fig:ref_image} shows all the six 
surges, kept under study, during their maximum phase as captured by three different filters namely, slit-jaw-images (SJI)/IRIS - 1330~{\AA}, 
AIA~171~{\AA}, and AIA~94~{\AA}. All surges were clearly visible in IRIS/SJI 1330~{\AA} filter (cf., top-row; Fig.~\ref{fig:ref_image}) which 
samples the cool plasma (i.e., Log T/K = 4.5). Signature of these surges, although less prominent, are also visible in AIA~171~{\AA} (cf., 
middle-row; Fig.~\ref{fig:ref_image}) which samples coronal emission (i.e., Log T/K = 5.8; \citealt{Lemen2012}). While, the hot temperature 
filter (i.e., AIA 94~{\AA} {--} Log T/K = 6.8; \citealt{Lemen2012}) gives very less signature of surges near their 
respective bases as displayed in the bottom-panel of Fig.~\ref{fig:ref_image}. 

\subsection{Evolution of Surges}

\begin{figure*}
    \mbox{
    \includegraphics[trim = 5.5cm 0.0cm 6.8cm 0.0cm,height=0.55\textheight, width=0.25\textwidth]{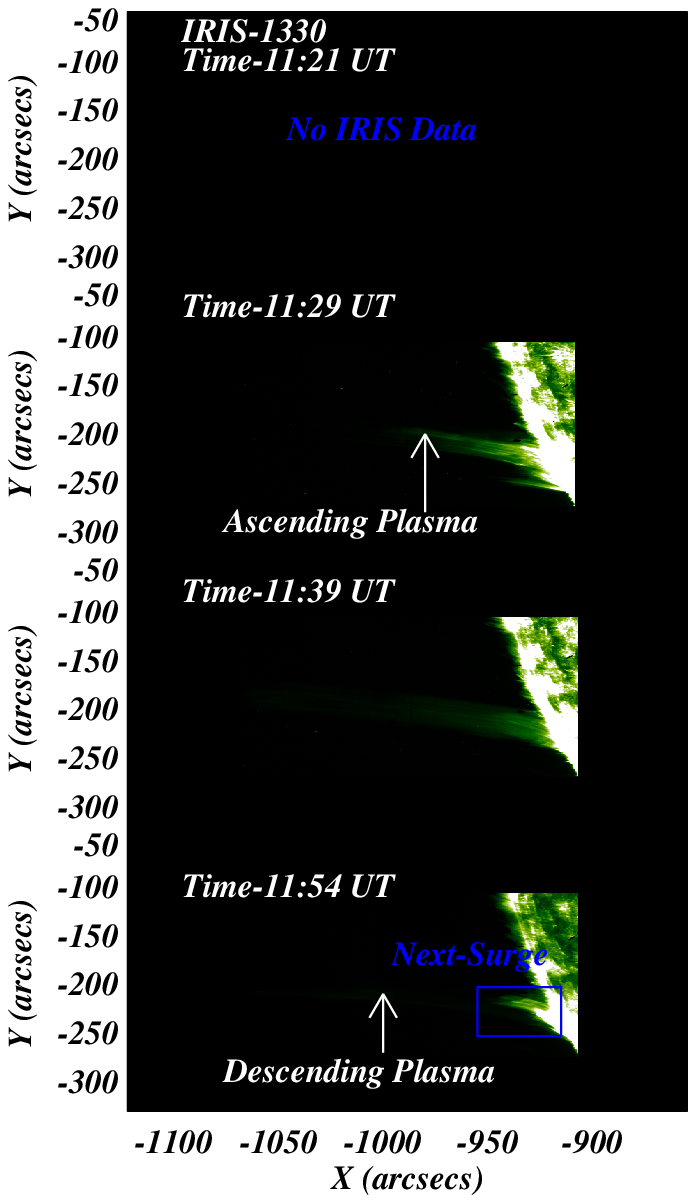}
    \includegraphics[trim = 5.5cm 0.0cm 6.8cm 0.0cm,height=0.55\textheight, width=0.25\textwidth]{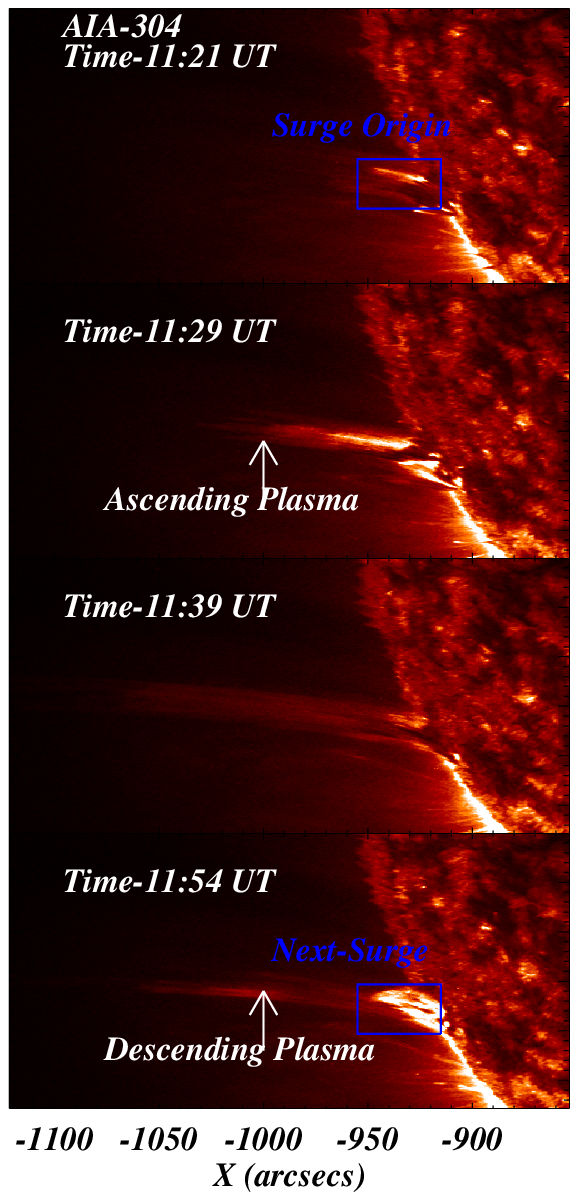}
    \includegraphics[trim = 5.5cm 0.0cm 6.8cm 0.0cm,height=0.55\textheight, width=0.25\textwidth]{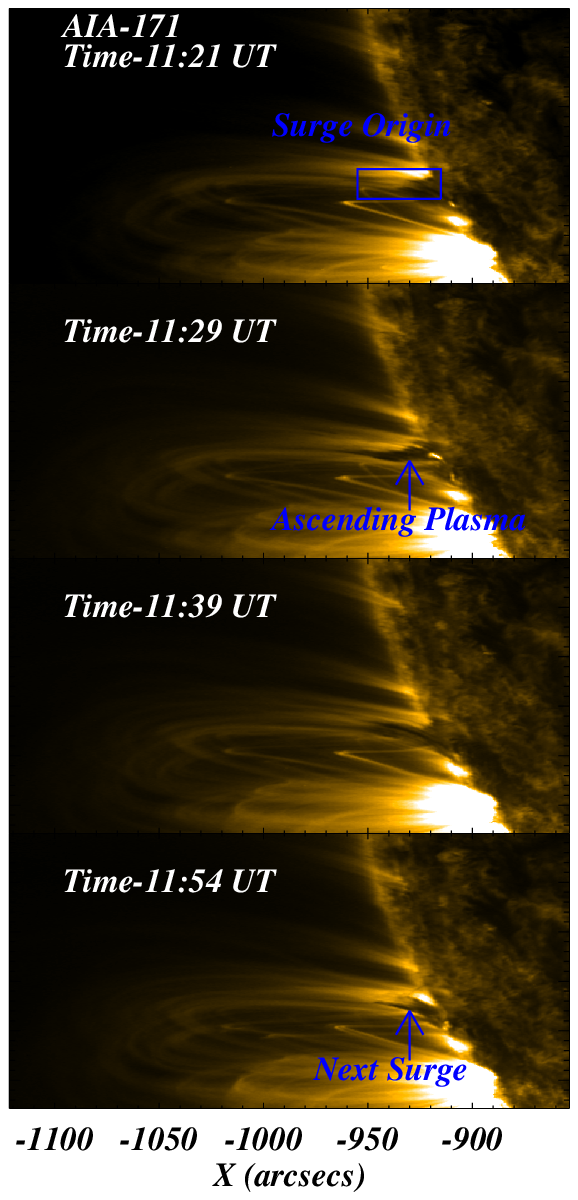}
    \includegraphics[trim = 5.5cm 0.0cm 6.8cm 0.0cm,height=0.55\textheight, width=0.25\textwidth]{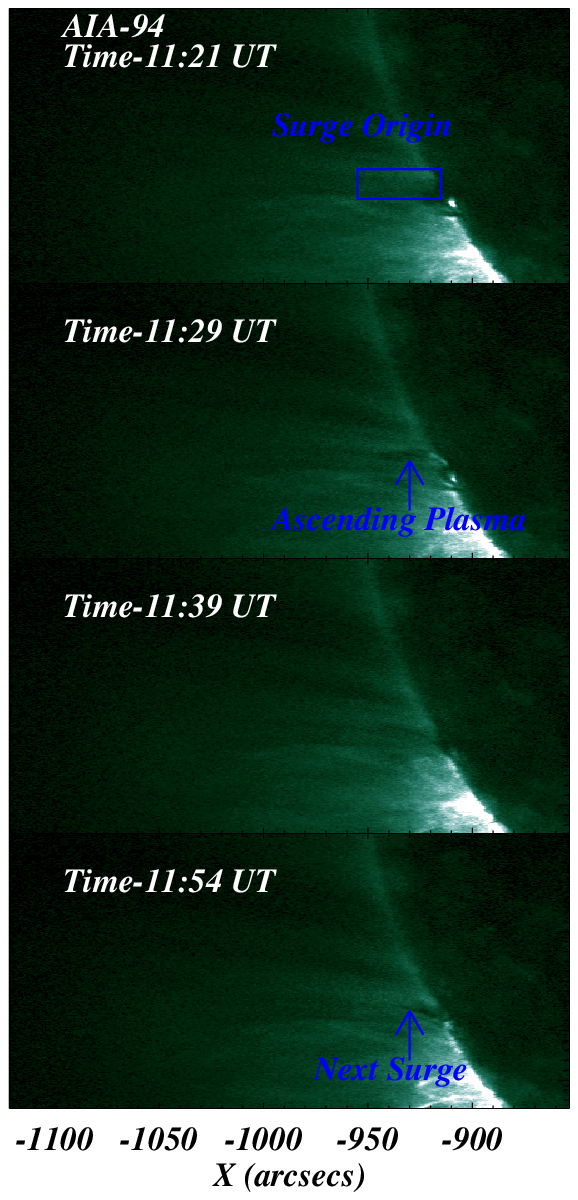}
    }
    \caption{The evolution of the first surge (S1) in IRIS 1330~{\AA} (first column), AIA 304~{\AA} (second column), AIA 171~{\AA} 
	(third column), and AIA 94~{\AA} (fourth column) is presented in this figure. We do see the usual behavior of the surge plasma, i.e., initially plasma flows up and later falls back onto the solar surface. The surge is clearly visible in IRIS 1330~{\AA} and AIA 304~{\AA}, while surge emission is comparatively less in AIA 171~{\AA}. Further, we see few traces (especially near the base) in the hot temperature filter of AIA (i.e., 94~{\AA}).}
    \label{fig:evol_surge}
\end{figure*}

In this section, we describe the evolution of the first surge, using IRIS/SJI 1330~{\AA}, AIA~304~{\AA}, AIA~171~{\AA}, and 
AIA~94{\AA} observations. This surge was originated around 11:21~UT, however, IRIS/SJI 1330~{\AA} observation is not aviliable at 
this particular time (top-row; Fig~\ref{fig:evol_surge}). The concerned surge is clearly visible in AIA~304{\AA}, however, 
AIA~171~{\AA} filter shows emissions only in and around the base of surge. Interestingly, we do see the less 
and faded signature of the surge in the hot-temperature filter (AIA~94~{\AA}; top-right most panel). The site of origin of the 
surge is outlined by blue boxes in AIA~304{\AA}, AIA~171{\AA}, and AIA~94{\AA} filters (see; top-row - Fig~\ref{fig:evol_surge}). 

At T = 11:29~UT, the surge is visible in all filters and the plasma is found to have an upward flow as shown by white 
arrows in all the four panels (see second-row; Fig.~\ref{fig:evol_surge}). Around T = 11:39~UT, the surge attains it's maximum phase. 
In the last panel (at time T = 11:54~UT), we do see the signature of falling plasma, back towards the solar surface. Here AIA~94~{\AA} observation shows 
the least emission as compared to other filters of AIA. In addition, it should be noted that AIA~171~{\AA} images 
clearly show the presence of closed loop structures and the surge is occuring around the one of the footpoints. Finally, 
we can say that this surge is showing standard plasma dynamics, i.e., first the plasma is shooting up, and then it is falling back 
towards the solar surface because of the solar gravity. The similar plasma dynamics is observed for other surges, kept under study, too. 

Meanwhile, we see the signature of origin of next surge in the last phase of this surge which is outlined by 
blue boxes in all four-panels of the last row of Fig.~\ref{fig:evol_surge}. Signature of the next surge is visible even 
in the very hot temperature filter (AIA~94~{\AA}; rightmost panel in last row of Fig.~\ref{fig:evol_surge}). 

\begin{figure*}
    \mbox{
    \includegraphics[trim = 3.5cm 0.0cm 4.5cm 0.0cm,scale=0.92]{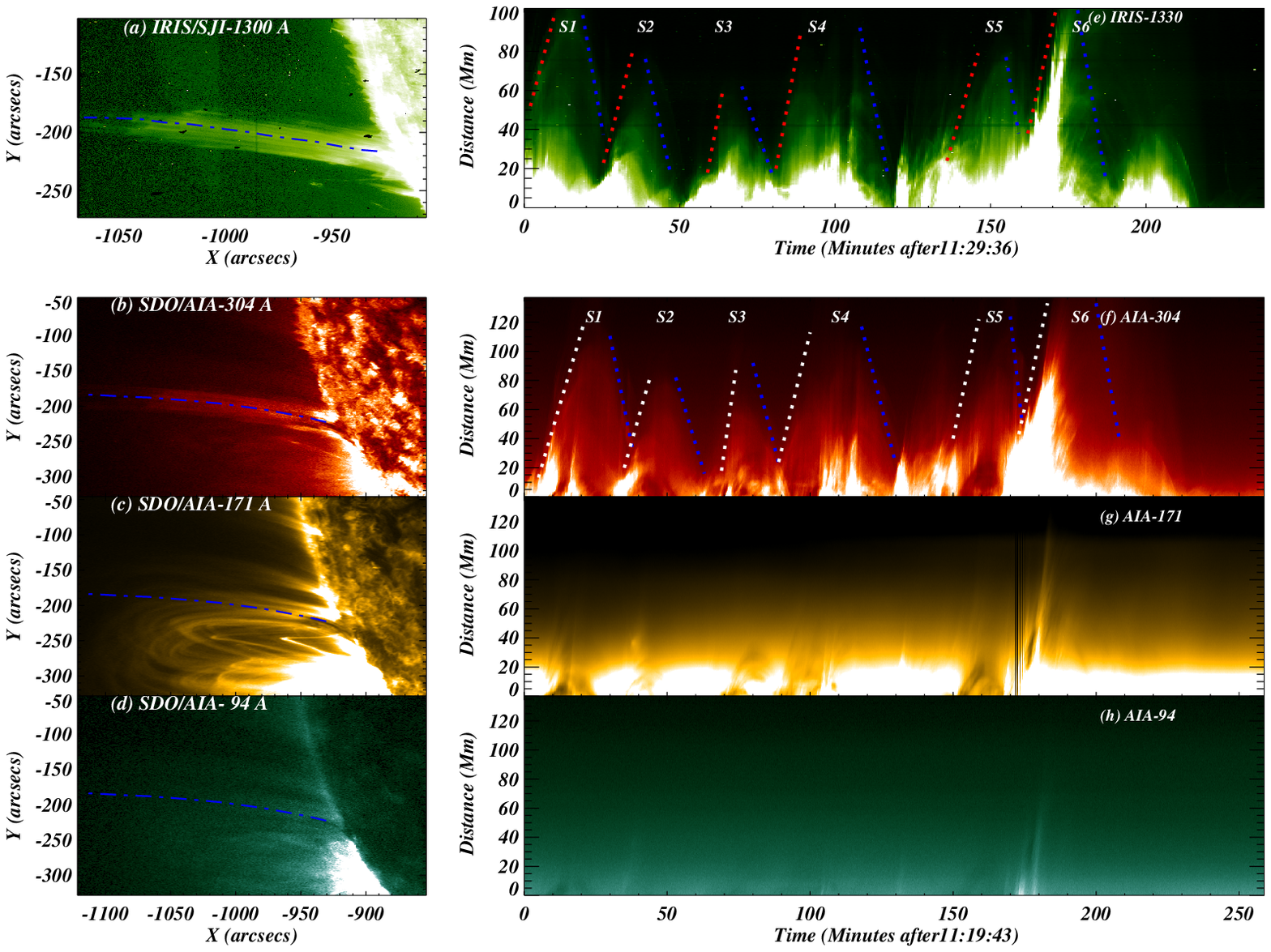}
    }
   \caption{Figure shows the selected slit by blue dash-dotted line, which is used to produce the height-time profiles, in IRIS/SJI 1330~{\AA}(panel a), AIA~304~{\AA} (panel b), 
           AIA~171~{\AA} (panel c), and AIA~94~{\AA} (panel d). While panel e, f, g, and h show the height-time profiles of IRIS 1330~{\AA}, AIA~304~{\AA}, AIA~171~{\AA}, 
           and AIA~94~{\AA}, respectively. We 
	have drawn slits during upflow (i.e., red and white dashed lines in panel e and f) and downflow phase (blue dashed 
	lines in panel e and f), as well. These slits were utilized to estimate various properties (e.g., up flow, 
	downflow speeds, acceleration, declaration, and life-times) of all six surges, observed in,  IRIS~1330~{\AA} (panel e) and AIA~304~{\AA} (panel f). These properties are tabulated in table~\ref{tab:surge_para}. We only see moderate signatures 
        of the surges in height-time profile of AIA~171~{\AA} and AIA~94~{\AA} (see; panel g and h), that's why the slits are not plotted here in.}
 \label{fig:ht}
\end{figure*}

Next, we estimated kinematic properties of these surges e.g., life-time, velocity, acceleration, and deceleration etc. Kinematics was 
estimated by applying height-time (ht) technique to each surge and thus, ht profiles have been created in all the filters. These
ht profiles are displayed in figure~\ref{fig:ht}. Left coulmn of Fig.\ref{fig:ht} shows the reference image from IRIS/SJI~1330~{\AA} 
(panel a), AIA~304~{\AA} (panel b), AIA~171~{\AA} (panel c), and AIA~94~{\AA} (panel d), and slit is also plotted on each reference image 
(see blue-dashed lines). Note that we have used this slit to produce the ht profile for IRIS/SJI~1330~{\AA} (panel d), AIA~171~{\AA} 
(panel e), and AIA~94~{\AA} (panel f).  

Evolution of all the surges, under study, is clearly visible in IRIS/SJI 1330~{\AA} and AIA~304~{\AA} filters. The ht profiles are depicted by 
S1, S2 and so on. Ascending and descending phase of each surge, in thse two filters, are overplotted with the dashed lines to calculate the 
upflow velocity, downflow velocity, acceleration, deceleration and life time. Ascending phases in IRIS-1330~{\AA} and AIA~304~{\AA} are oveplotted 
by dashed red and white lines, respectively. Whereas, descending phases for both the filters are overplotted by dashed blue lines. On the basis of drawn 
path on the ht profile, S1 starts from 11:19~UT and lasts till 11:54~UT with a total life-time of $\sim$37 minutes. In S1 plasma is ascending 
up with a veocity of $\sim$100.0 km s$^{-1}$, alongwith the acceleration of 208.33 m s$^{-2}$ and then plasma falls back with a speed of 155.0 km s$^{-1}$ 
along with deceleration of 393.80 m s$^{-2}$. In the same way, we have estimated these parameter for other surges too. Parameters estimated for all 
the surges (from IRIS~1330~{\AA} ht profiles) are tabulated in table~\ref{tab:surge_para}). Additionally, we would like to emphasize that ht profiles created 
from AIA~171~{\AA} and AIA~94~{\AA} also show the signatures of surges but less prominent in comparison to the IRIS/SJI 1330~{\AA} and AIA~304~{\AA}.

Please note that the kinematics of surges has already studied in many of previous works. The previous findings reported that 
surges move upward at with the speed of 20{--}200 km s$^{-1}$, reach heights of up to 200,000 km, and typically last for 10{--}20 minutes (\citealt{Roy1973,Bruzek1977,Chae1999}). 
Recently, \cite{Bong2014} have investigated 14 surges using Hinode/SOT observations, and they reported that the lifetime of the surges vary from 28.0 to 
91.0 minutes, with a mean lifetime of 45.0 minutes. Further, they have reported acceleration of 100 m s$^{-2}$ and deceleration of 70 m s$^{-2}$ in surges. 
\cite{Kayshap2013a} have also reported that surge plasma propagates with the speed of $\sim$100 km s$^{-1}$ while the lifetime of the surge is more 
than 20 minutes. Findings of the present study are in accordance with that of previous studies with reference to surge kinematics.
\begin{table*}
\begin{tabular}{|l|l|l|l|l|l|l|}
\hline
Parameters & S1 & S2 & S3 & S4 & S5 & S6 \\
\hline
Time-Span (UT) & 11:19-11:56 & 11:54-12:20 & 12:20-12:49 & 12:50 - 13:30 & 13:36-13:59 & 14:05-14:55\\
\hline
Life-time (Minutes) & $\sim$ 37.0 & $\sim$26.0 & $\sim$ 29.0 & $\sim$40.0 & $\sim$ 23.0 & $\sim$ 50.0\\
\hline
Upflows speed (km s$^{-1}$) & 100.00 & 101.56 & 143.33 & 145.83 & 91.66 & 116.66 \\
\hline
Acceleration (m s$^{-1}$) & 208.33 & 184.19 & 477.76 & 303.81 & 152.76 & 216.60 \\
\hline
downflows speed (km s$^{-1}$) & 155.0 & 120.83 & 76.66 & 140.74 & 162.52 & 159.26 \\
\hline
Deceleration (m s$^{-1}$) & 294.92 & 251.79 & 127.76 & 260.63 & 677.08 & 294.92 \\
\hline 
\end{tabular}

\caption{\label{tab:surge_para} Various parameters of studied surges (e.g., time, life-time, upflows, downflows, acceleration 
	and decelration) are tabulated in this table.}
\end{table*}
\subsection{Thermal Structures of Surges}

\begin{figure*}
    \mbox{
    \includegraphics[trim = 2.5cm 0.0cm 4.5cm 0.0cm,scale=0.92]{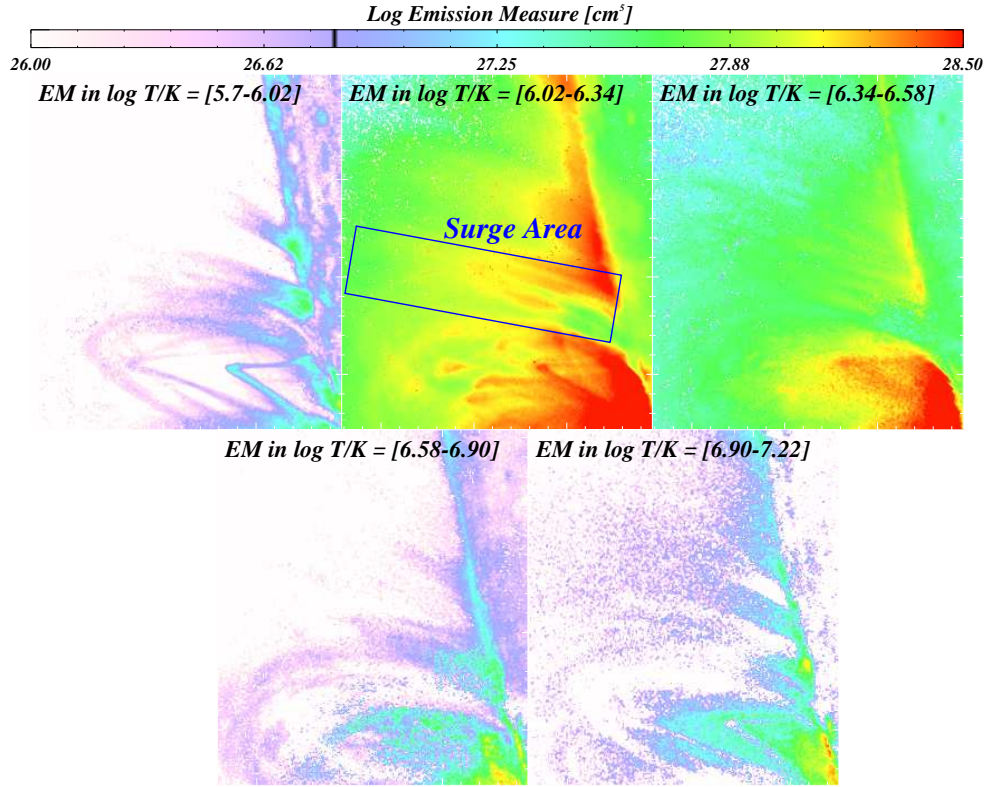}
  }
	\caption{In this figure, we present the EM maps in various temperature ranges (temperature range is written on each 
	EM map) for S1.  
        The blue rectangular box in the top-middle panel outlines the surge area, and a significant surge's 
	emission does exist in this temperature range (i.e; log T/K = 6.0 - 6.3). We do see a relatively small emission near 
	the base in the next temperature bin (i.e., right-top panel; log T/K = 6.3-6.6). Interestingly, in temperature bin of 
	log T/K = 6.6-6.9, a significant emission especially near the base justifies the heated base. Again, the next higher 
	temperature bin (i.e., log T/k = 6.9-7.2) is showing very little emission near the base.} 
 \label{fig:s1_dem}
\end{figure*}

We investigated the thermal nature of surges with the help of SDO/AIA observations. We have performed the DEM analysis using 
the sparse inversion method (\citealt{Cheung2015}) for six different filters of SDO/AIA (viz. 94~{\AA}, 131~{\AA}, 171~{\AA}, 
193~{\AA}, 211~{\AA}, and 335~{\AA}). The sparse inversion has been validated against a diverse range of
thermal model, and it gives positive definite DEM solutions. In addition, the sparse diversion technique can be used in the 
generation of a sequence of DEM images as it is a very fast method (e.g., \citealt{Cheung2015, Jing2017}). DEM solution was 
obtained on a pixel-by-pixel basis on a temperature bin of 0.04 (log-scale) within the range of log T/K = 5.7 to 7.78. DEM estimated,thus, 
for each pixel lies within our region-of-interest (ROI). 
Figure~\ref{fig:s1_dem} shows EM maps from a sparse inversion of AIA observations of NOAA AR 12114 along with S1 during its 
maximum phase. EM (i.e., total DEM emission for particular temperature interval) is shown in the different temperature bins. 
For instance, the top-left panel shows EM in the temperature regime of log T/K = [5.7 - 6.0]. Here, a little emission can be 
seen near the base of the surge. While the top middle-panel shows the EM for a temperature range log T/K = [6.0 - 6.3]. The 
surge is clearly visible within this temperature bin, as outlined by a blue rectangular box (see; 
top-middle panel). Spire of the surge has major emission in this particular temperature bin. Although the base of surge 
also emits significantly, but it was little in comparison to emission from spire region. In the next temperature bin (i.e, log T/K = 
[6.3 - 6.6]), we do not see much emission from the surge. Moderate emission is visible near the base of the surge in a 
small area in this temperature bin (see; the top-right panel of Fig.~\ref{fig:s1_dem}). However, a major emission around the base of 
the surge is visible in the temperature bin of log T/K = [6.6 - 6.9] (see bottom left-panel of Fig.~\ref{fig:s1_dem}). We also see 
little emission near the base of the surge in the temperature bin of log T/K = [6.9 - 7.2]. While investigating other surges 
(i.e., from S2 to S6) using the similar approach we found similar behavior (plots not shown here).

\begin{figure*}
    \mbox{
    \includegraphics[trim = 1.5cm 0.0cm 1.5cm 1.0cm,scale=0.80]{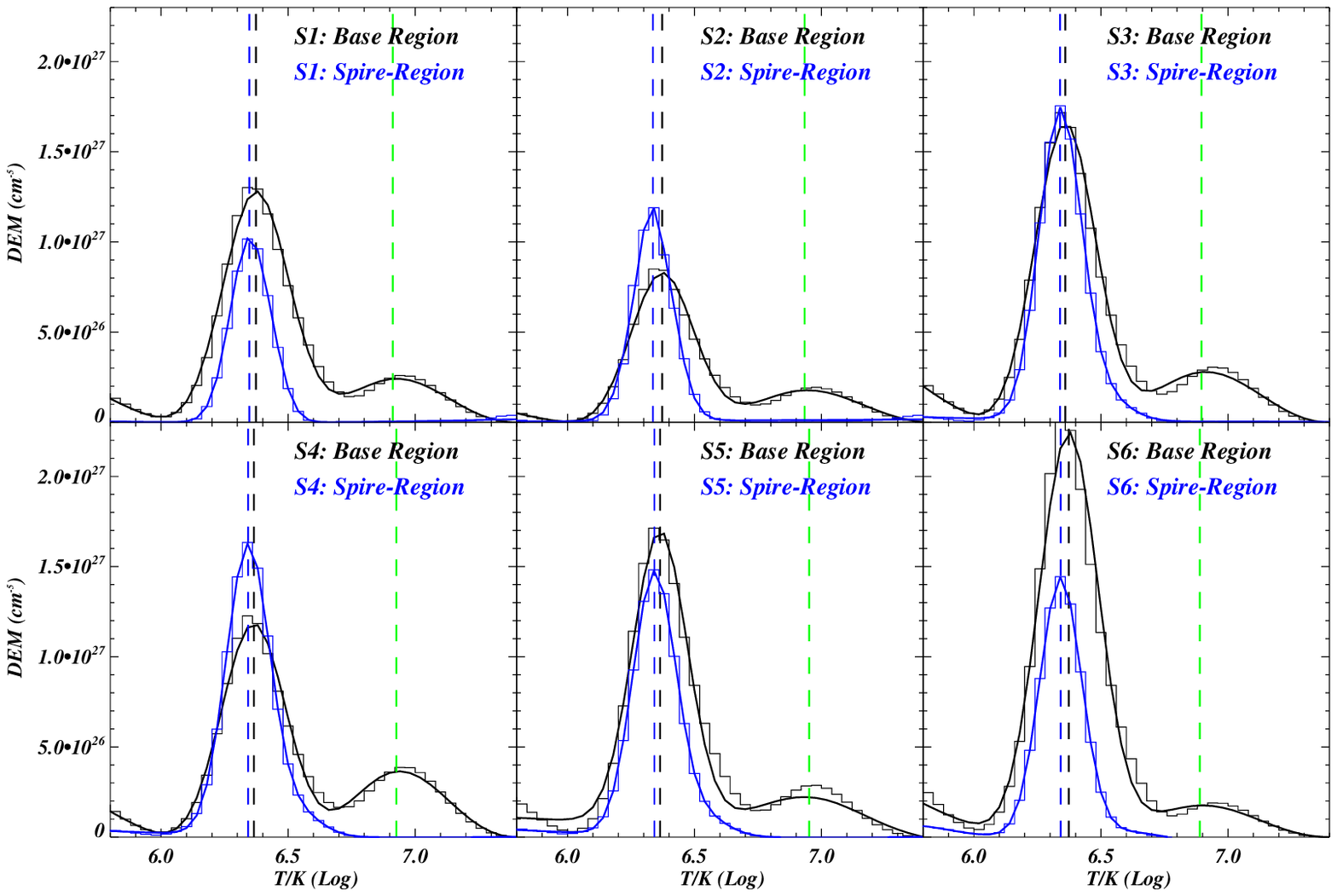}
    }
   \caption{The figure shows the distribuation of DEM from the base (black histogram) and spire (blue histogram) for all six 
	surges. Displayed DEM maps are obtained from the maximum phase of the corresponding surge.
        DEM distribution of base region shows two peaks. While, DEM distribition of spire has only one peak. Hence, we 
	have fitted the base DEM with double Gaussian function (black curve) and spire DEM with a single Gaussian functions
	(blue curve).}
        
 \label{fig:dem_base_spire}
\end{figure*}

Further, we investigated the DEM distribution from two different regions (i.e., base and spire region) of all the surges under study. 
The top-left column of figure~\ref{fig:dem_base_spire} displays the DEM from base (black histogram) and spire (blue histogram) of S1. 
In this figure, two peaks for base region at log T/K = 6.37 and log T/K = 6.93 are clearly visible. This DEM data is fitted with 
double Gaussian (see; overplotted black solid line {--} top-left panel of Fig.~\ref{fig:dem_base_spire}), justifying the presence of 
two separate distributions. Peaks of these two Gaussians are shown by vertical black-(cool component) and green-(hot component) dashed 
lines. It should be noted that the cool component (i.e., around log T/K = 6.37) has a higher DEM compared to the DEM
of hot peak. And, the spread (i.e., Gaussian sigma) of the hot component DEM distribution is almost double (0.341) of that for the 
cool component DEM distribution (0.189). As the DEM of spire shows only a single peak, it was fitted with a single Gaussian (see overplotted 
solid blue line). The dashed blue vertical line shows the peak of DEM (at  log T/K = 6.34). Thus, we can say that hot component does not 
occur within the spire plasma of S1. In addition, the spread of DEM distribuation of the cool component is found 
narrower for spire (0.11) as compared to that of base (0.189). Similar analysis have been performed for other surges as well (S2: 
top-middle panel, S3: top-right panel, S4: bottom-left panel, S5: bottom-middle panel, 
and S6: bottom-right panel) and it is observed that all the surges show similar behavior. Parameters obtained from the Gaussian fitting of DEM distributions 
are tabulated in table~\ref{tab:surge_base_dem} for base and in table~\ref{tab:surge_spire_dem} for spire, respectively. Tabulated parameters 
suggest that: (a) there is comparatively higher cool emission at the base of surges, although, the hot emission is also significant 
at the base of each surge. (b) peak position for cool component is around log T/K = 6.35 and that of hot component is around log T/K = 6.95 (c) spread of 
the hot component is double as that of cool component, and (d) spread of the DEM distribution for cool component is narrow for spire region. 

\begin{table*}
\begin{tabular}{|l|l|l|l|}
\hline
Surge id (Base)          & DEM Peak value[1/2] (cm$^{-5})$)                              & Centroid temperature[1/2] (log T/K)  & Gaussian Sigma[1/2] (logT) \\
\hline
S1                 & 1.4$\times$10$^{27}$/4.4$\times$10$^{26}$   & 6.37/6.91              & 0.186/0.341\\
\hline
S2                 & 8.8$\times$10$^{26}$/2.8$\times$10$^{26}$   & 6.37/6.93              & 0.177/0.344\\
\hline
S3                 & 1.78$\times$10$^{27}$/5.57$\times$10$^{26}$ & 6.35/6.89              & 0.171/0.371\\ 
\hline 
S4                 & 1.34$\times$10$^{27}$/5.92$\times$10$^{26}$ & 6.36/6.62              & 0.184/0.311\\
\hline
S5                 & 1.62$\times$10$^{27}$/2.03$\times$10$^{26}$ & 6.36/6.95              & 0.152/0.305\\
\hline
S6                 & 2.29$\times$10$^{27}$/3.44$\times$10$^{26}$ & 6.37/6.88              & 0.166/0.377\\
\hline
\end{tabular}

\caption{\label{tab:surge_base_dem} We fitted the double Gaussian function on the base DEM of each surge. 
The Gaussian fitting gives the peak values of DEM, centroid temperature, and Gaussian width (spread of the distribution) of 
cool component and hot component. We tabulated all three parameters for the cool component (with suffix 1) as well as the 
hot component (with suffix 2) in this table. "1/2" referes to "cool component/hot component".}
\end{table*}

\begin{table*}
\begin{tabular}{|l|l|l|l|}
\hline
Surge id (spire)    & DEM Peak value(cm$^{-5})$)               & Centroid temperature (log T/K)  & Gaussian Sigma (log T/K) \\
\hline
S1                 & 1.0$\times$10$^{27}$  & 6.34              & 0.11 \\
\hline
S2                 & 1.10$\times$10$^{27}$ & 6.34              & 0.11\\
\hline
S3                 & 1.71$\times$10$^{26}$ & 6.33              & 0.117\\
\hline
S4                 & 1.61$\times$10$^{27}$ & 6.34              & 0.121\\
\hline
S5                 & 1.45$\times$10$^{27}$ & 6.36              & 0.152\\
\hline
S6                 & 1.41$\times$10$^{27}$ & 6.34              & 0.115\\ 
\hline 
\end{tabular}

\caption{\label{tab:surge_spire_dem} Same as table~\ref{tab:surge_base_dem} but for the spire of surges. It should be noted 
        that spire of surges has only one peak unlike the DEM of base of the surges.}
\end{table*}

\subsection{Spectoscopic Evolution of Surges}

\begin{figure*}
    \mbox{
    \includegraphics[trim = 2.5cm 0.0cm 4.5cm 2.0cm,scale=0.92]{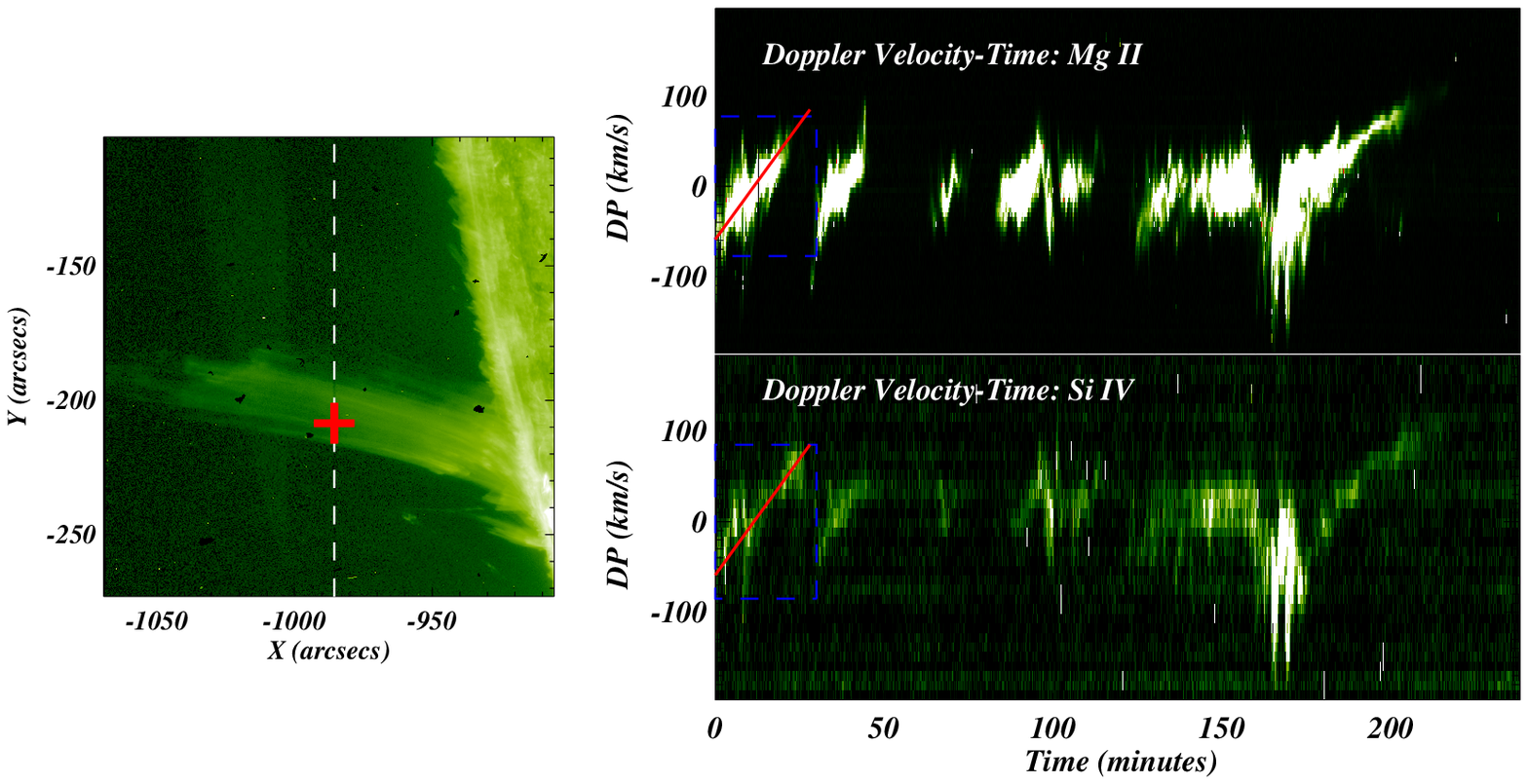}
    }
   \caption{Left-panel shows the IRIS/SJI~1330~{\AA} image of the surge along with the slit position (white-dashed line). 
	   The right-panel shows Doppler velocity-time diagram for Mg~{\sc ii} k (top-panel) and Si~{\sc iv} 1402.77~{\AA} 
	   (bottom-panel) that is taken from one particular y-position (i.e., y = -208.56 arcsec) shown by a plus red symbol in 
           the left-panel. The wavelength is transformed into Doppler velocity using standard rest wavelengths taken from CHIANTI atomic database, 
	   i.e., 1402.77~{\AA} and  2796.35~{\AA} for Si~{\sc iv} and Mg~{\sc ii} k, respectively. Please note that both 
	   spectral lines have significant emission during these surges as the counts are significantly high. First surge 
	   is outlined by the blue rectangular box in both panels. Initially, both lines are blueshifted that shifts towards 
	   the red-shifts with time. The transformation from blue- to red-shifts is very gentle as outlined by solid red 
	   lines in both panels.}
 \label{fig:surge_spectra}
\end{figure*}

IRIS provides high-resolution spectral observations of these  six homologous surges. IRIS slit has been 
placed in the spire region (i.e., far away from surge's bases; see white-dashed line in left-panel of 
Fig.~\ref{fig:surge_spectra}) of all the surges. Unfortunately, we do not have the spectroscopic observations 
near the base of these surges. However, the available IRIS spectroscopic observations (i.e, spectra far-away 
from the bases of surges) still provides important information about these surges.

We utilized few important IRIS lines, viz.  Mg~{\sc ii} k 2796.35~{\AA}, Mg~{\sc ii} h 2803.52~{\AA}, Si~{\sc iv} 1402.77~{\AA}, 
and O~{\sc iv} 1401.15~{\AA}, for the analysis of these surges. Left-panel of figure~\ref{fig:surge_spectra} displays the 
IRIS/SJI~1330~{\AA} image along with slit-position shown by white-dashed line, i.e., the location within which IRIS has captured the 
surge spectra. Further, the Doppler velocity-time (v-t) diagram for prominent Mg~{\sc ii} 2796.35~{\AA} (top-right panel) and Si~{\sc iv} 
1402.77~{\AA} (bottom-right panel) emissions are also displayed in the figure~\ref{fig:surge_spectra}. The wavelength is converted into 
Doppler velocity using the standard wavelength of these lines as provided by CHIANTI atomic database, namely, 
2796.35~{\AA} for Mg~{\sc ii} k and 1402.77~{\AA} for Si~{\sc iv} line. Signal-to-noise (SNR) ratio is very 
high during the surges, and the rest of the slit area is having almost no signals. All the surges are very clear in the v-t 
diagram of Mg~{\sc ii} 2796.35~{\AA} and Si~{\sc iv} 1402.77~{\AA} lines, i.e., bright areas in both right-panels of 
figure~\ref{fig:surge_spectra}.

Further, the v-t diagram also reveals that both the lines (i.e., Mg~{\sc ii} and Si~{\sc iv}) are  highly blueshifted at the time of 
origin. The blueshift decreases with time and inverts into redshift. Solid red lines on both v-t diagrams (i.e., S1; both right-panels; 
figure~\ref{fig:surge_spectra}) exhibit the linear transition for the doppler velocity. This transition was observed for the
other sugers too. In addition, the maximum blue- and red-shift is observed in the strongest surge (i.e. S6) 
which starts almost after 160.0 minutes from the start of first surge.

Both the lines were fitted with a single Gaussian to produce the Intensity, Doppler Velocity and Sigma Maps for complete observation 
which are shown in the appendix~\ref{append:maps} (see, figure~\ref{fig:mg_map_append} for Mg~{\sc ii} k 2796.35~{\AA} and 
figure~\ref{fig:siv_map_append} for Si~{\sc iv} 1402.77~{\AA} lines).

With the careful inspection of these spectroscopic parameter maps, following can be concluded about 
the general behavior of these surges: (1) blueshift dominates the first half lifetime of the surge, 
while redshift is dominent in the second half lifetime, (2) Gaussian sigma is higher for blueshift regions (i.e., first 
half lifetime of surges), and (3) Si~{\sc iv} line maps are fuzzier compared to Mg~{\sc ii} 
2796.35~{\AA}.

\subsubsection{Evolution of Si~{\sc iv}} 
\label{section:si_evol}

\begin{figure*}
    \mbox{
    \includegraphics[trim = 0.0cm 0.0cm 0.5cm 0.0cm,scale=0.85]{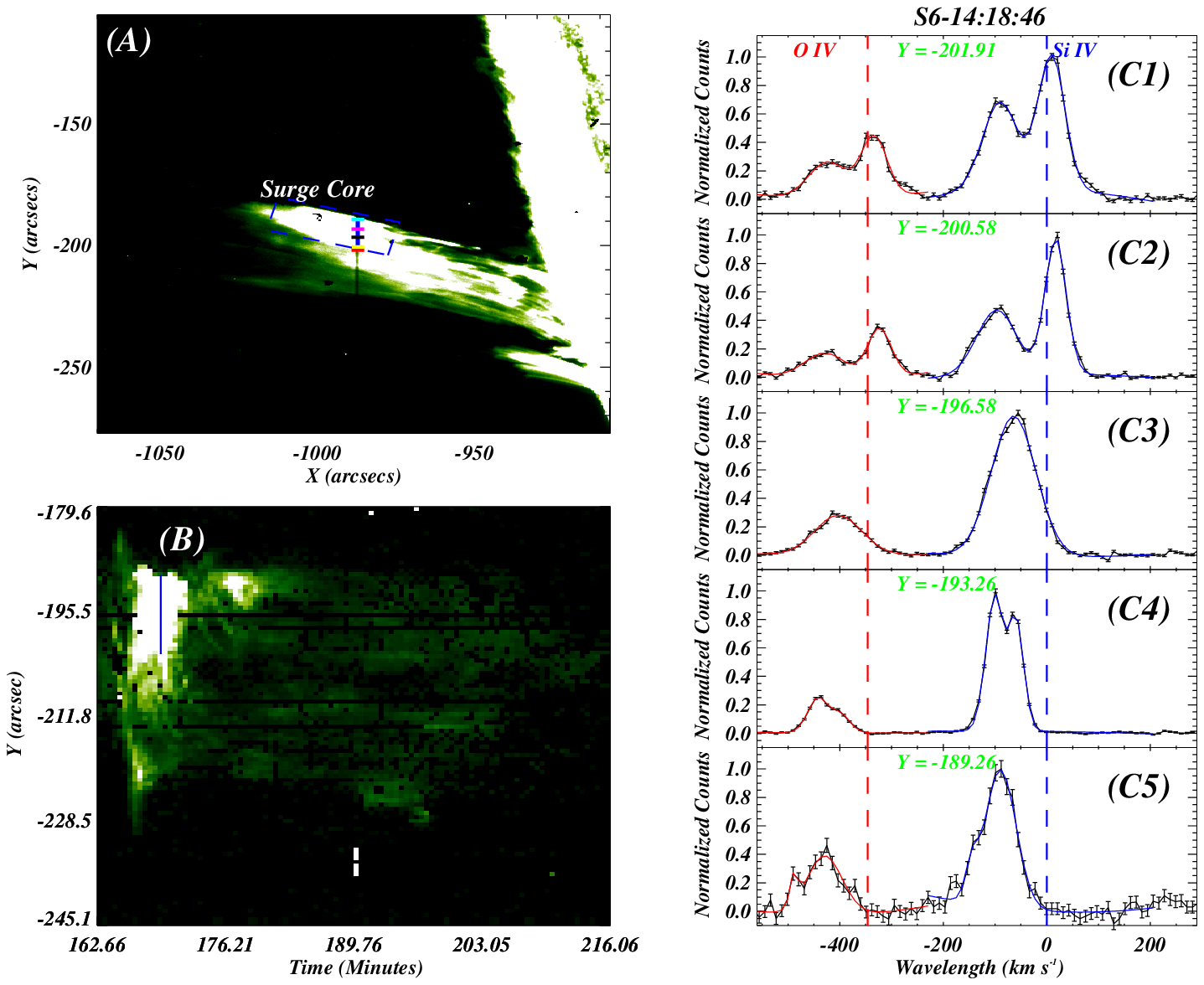}
    }

  \mbox{
    \includegraphics[trim = 1.90cm 0.0cm 0.5cm 0.0cm,scale=0.70]{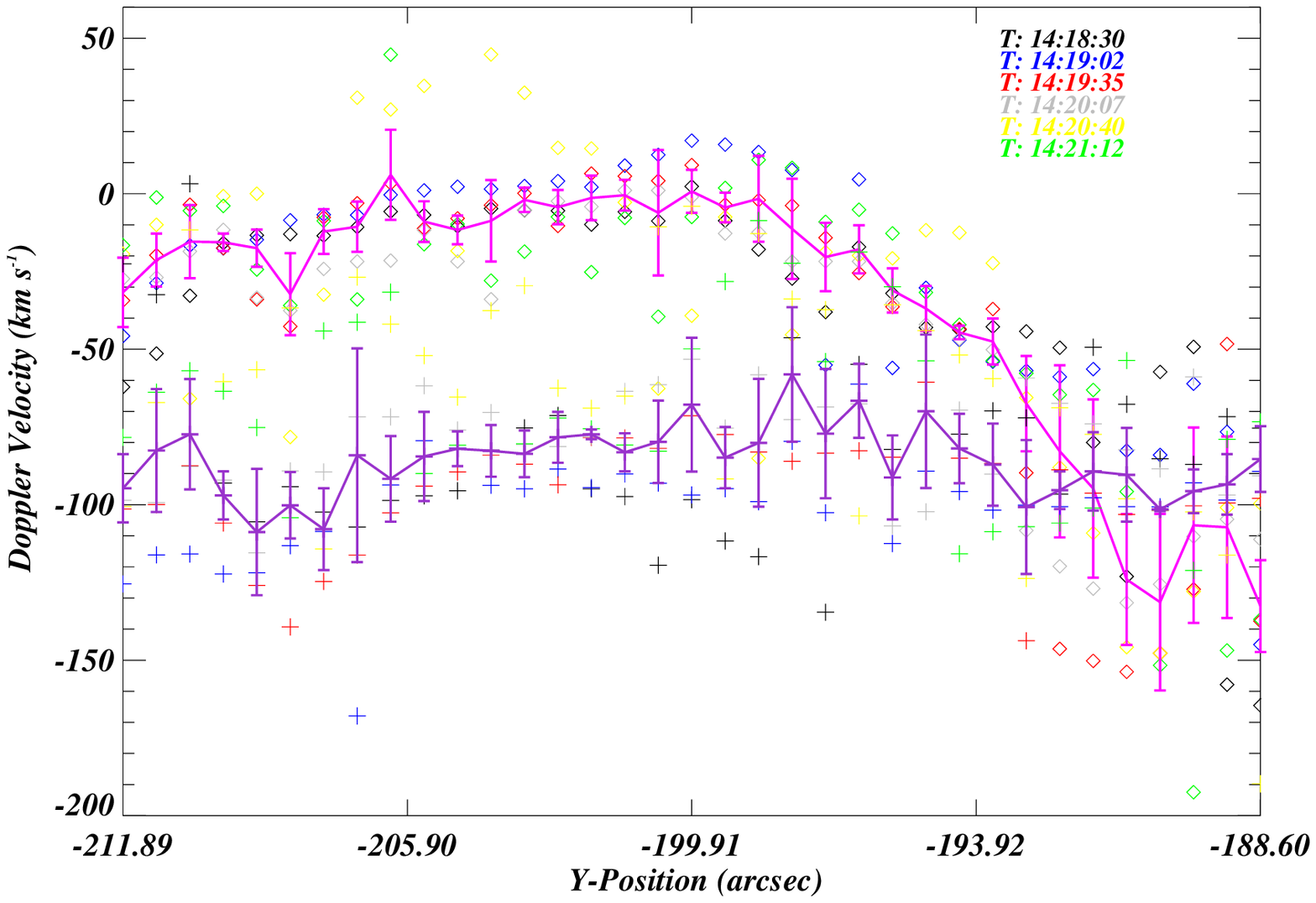}
    }

   \caption{Panel A displays IRIS/SJI 1330~{\AA} along with a blue-dashed rectangular box that outlines the surge's core plasma. 
	The panel B shows Si~{\sc iv} intensity-time diagram of S6. To show the spectral profiles, we selected five different 
        locations shown by horizontal slits of different colors in panel A, namely, C1-red, C2-yellow, C3-black, C4-magenta, and C5-cyan. 
        The panel C1 clearly shows the double peak in Si~{\sc iv} and O~{\sc iv} spectral profiles at the lower edge of S6 (i.e., Y = -201.99$"$). 
        Therefore, they are fitted with double Gaussian function (i.e., solid blue-line for Si~{\sc iv} and red-line for O~{\sc iv}). At the 
        next position in S6 (i.e., Y = -200.58$"$), we found the similar profiles (see; panel C2) while almost single Gaussian profiles exist for 
	both lines at Y = -196.58$"$ (see; panel C3). Further, at the next y-position = -193.26$"$ the double peak exists again in both lines, 
        however, both peaks are completely shifted into blue-regime. Finally, the similar behavior of both lines 
	is present at the other (upper) edge of S6 (panel C5). In the bottom panel, we have shown the variations of Doppler 
	velocities for both components of Si~{\sc iv} (i.e., both peaks) from six different times as mentioned in the bottom-panel. 
	The Doppler velocity of varying (non-constant) peak is shown by the diamond of different colors as per time while the Doppler velocity of 
        constant peak by plus signs of different colors as per the time. Finally, we averaged all time points at each y-position for both 
	peak and shown the averaged Doppler velocity by solid pink line (one peak) and purple line (another peak). We see that one 
	peak (solid pink line) shifts from red to blueshift as we move from one edge to another edge of S6, i.e., varying or non-constant component while another peak (solid purple line) shows almost constant Doppler velocity, i.e., constant component.}
 \label{fig:hm_one}
\end{figure*}

We have studied the evolution of Si~{\sc iv} 1402.77~{\AA} during the most dynamic surge (i.e., last surge: S6). IRIS/SJI 1330~{\AA} image of 
surge is displayed in the panel A of figure~\ref{fig:hm_one}. Blue-dashed rectangular box, along with a vertical solid blue line, 
outlines the core material of surge. The intensity-time map of Si~{\sc iv} for S6 is presented in panel B of figure~\ref{fig:hm_one}. 
A blue vertical line in the vicinity of surge's core plasma is drawn on  the Si~{\sc iv} intensity-time map (cf., panel B of 
figure~\ref{fig:hm_one}). This location has been used to study the evolution of Si~{\sc iv} spectral line. More precisely, 
thick horizontal dashes are displayed by different colors within the surge's core material (i.e., blue-dashed box in panel A) corresponding to 
the spectral profiles that are shown in the right column of figure~\ref{fig:hm_one} (i.e., panels C1 to C5). It must be noted that these locations 
cover the full vertical extent of this surge.

Panel C1 shows the Si~{\sc iv} and O~{\sc iv} spectral profiles from the bottom-end of surge's core plasma (i.e., 
red dash in blue-box; panel A). Si~{\sc iv} line shows two components, one on red-side and another on blue-side. Similar 
behavior is observed for O~{\sc iv} line, as well. Si~{\sc iv} and O~{\sc iv} lines were fitted with two Gaussians. Fitted curves 
are overplotted on the observed spectra by blue color for Si~{\sc iv} and red color for O~{\sc iv} spectral lines, respectively. 
Observed profiles are very well characterized using this double-Gaussian approach. Blue- and red-dashed vertical lines lie at 
the zero velocity for Si~{\sc iv} and O~{\sc iv}, respectively. 

Next-panel (C2), shows the profile from a location shown by yellow dash in the blue-rectangular box 
(see; panel A). Here, line profile exhibits a similar behaviour as that in C1. Panel C3 shows the profile obtained from near 
the middle of the core plasma (i.e., black dash in panel A). For both the lines, obtained profiles are having almost 
a single peak. Thus, these  profiles were fitted using a single Gaussian function. Further, as we move towards the upper edge of 
surge plasma, double peak profile starts appearing again for both the lines (see; panel C4{--}profile 
from the location shown by magenta dash in panel A and panel C5{--}profile from upper edge of surge plasma and it is shown by cyan dash in 
panel A). However, new peaks (or components) are blue-shifted for both lines. Interestingly, One component (or peak)  
of Si~{\sc iv} is always constant at -100 km/s at this particular instant of time (shown in panels from C1 to C5) that is defined as 
the constant velocity component.
On the contrary, other component (or peak) of Si~{\sc iv} changes from red-side (near the bottom edge) to blue-side 
(near the top-edge) along with the absence of this particular component (i.e., single Gaussian{--} constant component 
at -100 km/s) around the middle of S6. This secondary component, which changes from redshifts (at bottom edge) to 
blueshifts (at upper edge), is defined as the varying (or non-constant velocity) component. This finding refers to the observation at 
a single instant of time (i.e. 14:18:46), for different positions across the surge.  

We have chosen six different times that are mentioned in the bottom panel of Fig.~\ref{fig:hm_one}. We estimated the Doppler 
velocities of constant and varying (non-constant) components (as defined above) for all the Y-positions across S6 at the 
mentioned six different times. Deduced Doppler velocities, as a function of Y-position (i.e., positions across S6), are shown in the bottom panel of Fig.~\ref{fig:hm_one}. 
Doppler velocites of constant and non-constant components are shown by plus and diamond symbols, respectively. Next, we 
averaged all the Doppler velocites at a given Y-position (for all the considered time instants) for both the constant and non-constant 
components, separately. Averaged Doppler velocity profiles are overplotted by solid pink and purple lines for constant and non-constant
components, respectively. Here, error bars refer to the standard deviations at each Y-position. 
From this analysis, it can be concluded that Si~{\sc iv} spectral line has constant (i.e., having 
Doppler velocities around -100 km s$^{-1}$ at all y positions during the selected times), and non-constant component. Doppler velocity of this 
non-constant component changes from low red-shifts of $\sim$20 km s$^{-1}$ (at the bottom edge of S6) to very high blue shifts of $\sim$-140.0 
km s$^{-1}$ (at the top edge of S6). We performed the similar analysis for other surges too (i.e., for S1 to S5) and a similar 
behaviour of Si~{\sc iv} spectral line profiles for rest of the surges is found (see appendix~\ref{append:evol_profile}).

\subsubsection{Opacity $\&$ Evolution of Mg~{\sc ii} resonance lines in the homologous surges}
\label{sect:mg_opacity}
Mg~{\sc ii} k $\&$ h lines are transitions to a common lower level from finely splitted upper levels (\citealt{Leen2013, Kerr2015}). 
Intensity ratio of Mg~{\sc ii} k $\&$ h (i.e., R$_{kh}$) lines can be used to investigate the opacity (e.g., \citealt{Sch1997,Math1999}). 
The transitions of Mg~{\sc ii} resonance lines involve the same element (Mg) in the same ionization state. The escape probability of a 
photon is unity for optically thin conditions. Hence, in optically thin conditions, intensity ratio of the Mg~{\sc ii} k 2796.35~{\AA}
to Mg~{\sc ii} h 2803.52~{\AA} is simply the ratio of the oscillator strengths, i.e., R$_{kh}$ should be equal or greater than two 
(\citealt{Kerr2015}). While, in case of optically thick atmosphere, this ratio is smaller (\citealt{Kerr2015}).
\begin{figure*}
    \mbox{
 \hspace{-2.0cm}
    \includegraphics[trim = -1.0cm 0.0cm 0.5cm 0.0cm,scale=0.80]{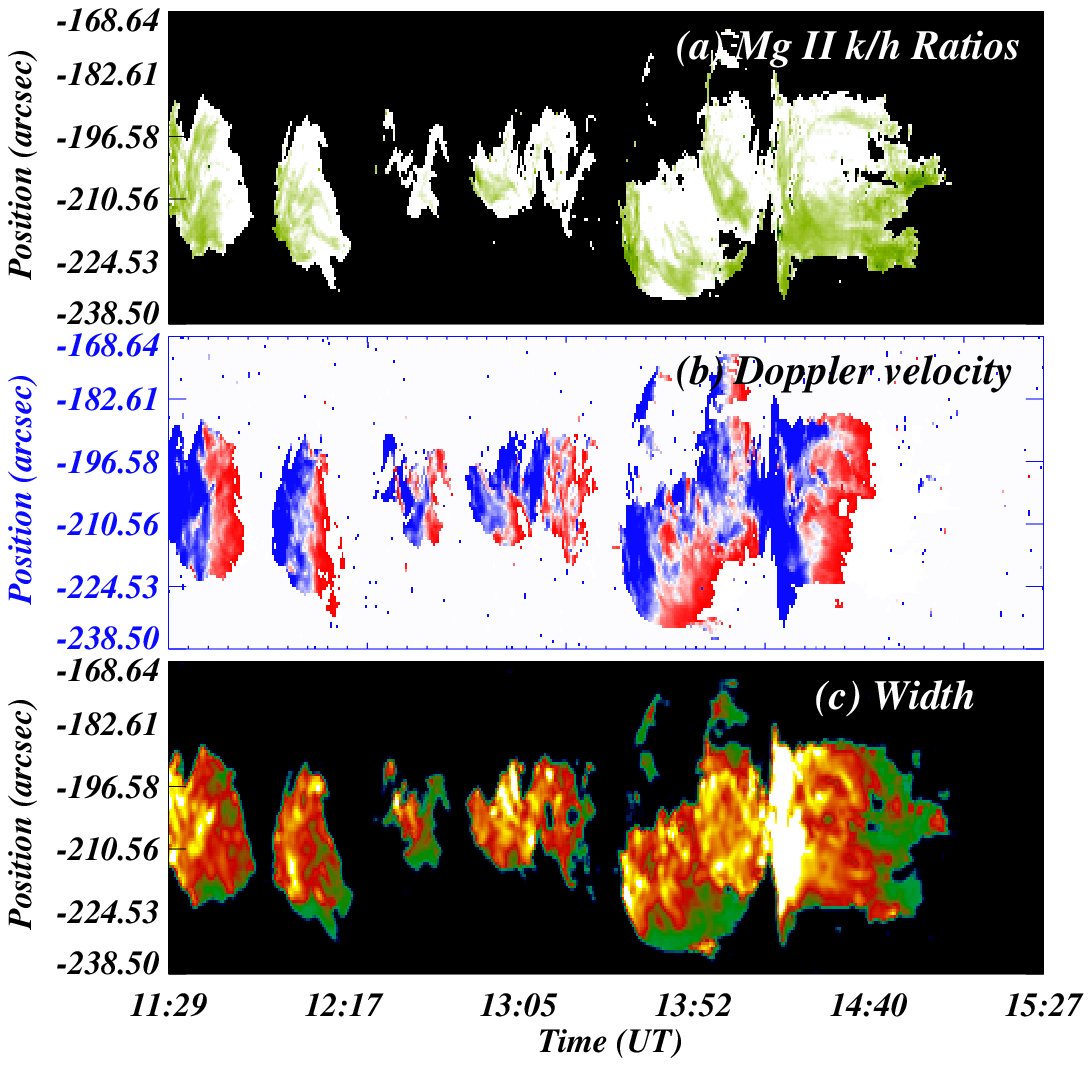}
}
\mbox{
    \includegraphics[trim = 2.0cm 0.0cm 0.5cm 2.0cm,scale=0.80]{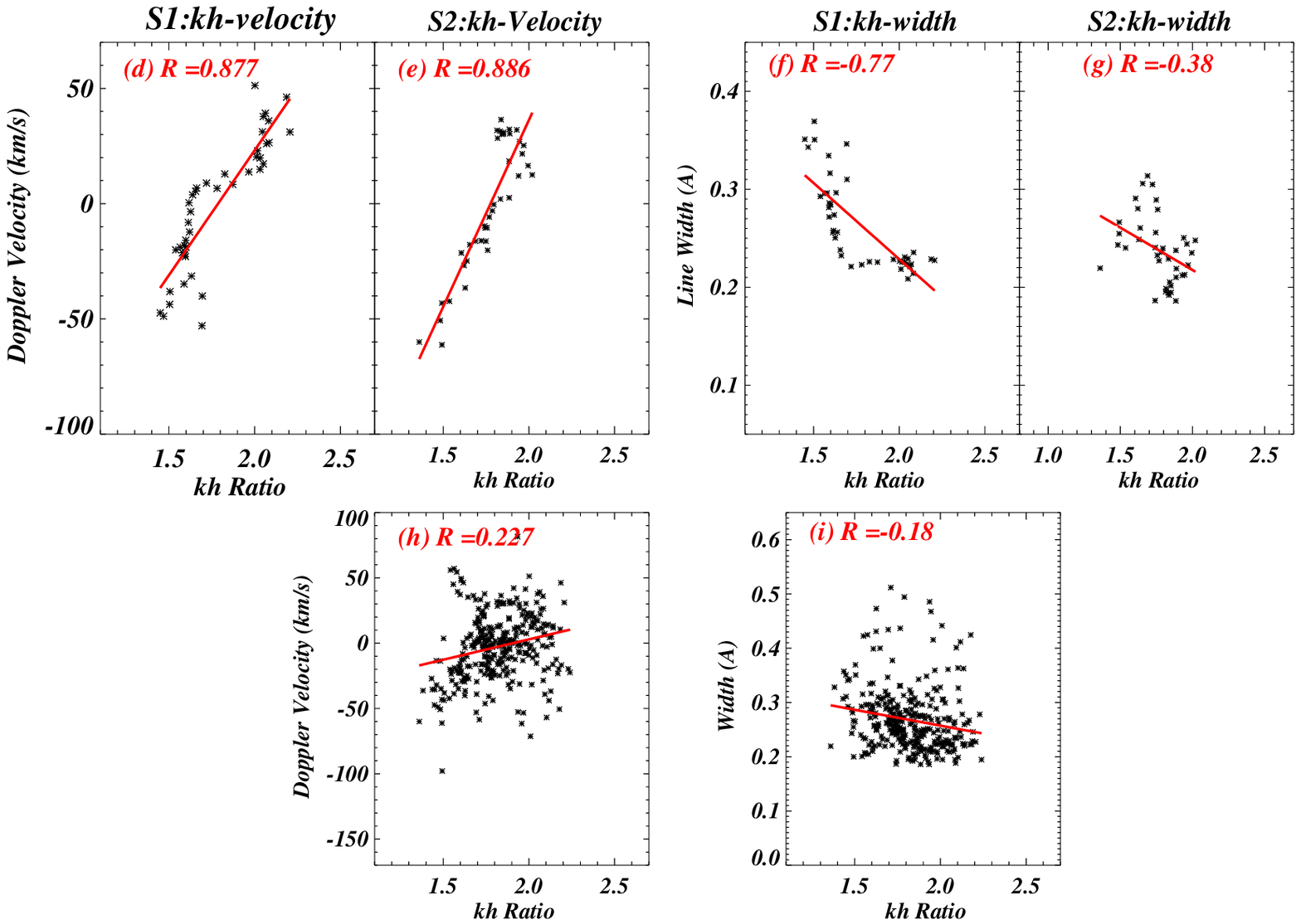}\
}
   \caption{In this figure, we have shown three maps, namely, Mg~{\sc ii} k/h ratio map (R$_{kh}$; panel a), Doppler velocity map 
	(panel b), and width map (panel c). Further, we have shown the correlations, namely, (1) correlation between 
	Doppler velocity and R$_{kh}$ for S1 (panel d) and S2 (panel e) and (2) correlation between Gaussian width and R$_{kh}$ for 
        S1 (panel f) and S2 (panel g). We found the tight and positive correlation between Doppler velocity and R$_{kh}$ for S1 (Pearson 
        coefficient = 0.877) and S2 ((Pearson coefficient = 0.886). While we found a negative correlation between Gaussian width and
        R$_{kh}$ for S1 (Pearson coefficient = -0.77) and a weak negative correlation between Gaussian width and R$_{kh}$ for S2 (Pearson 
        coefficient = -0.38). We collected these three parameters (i.e., R$_{kh}$, Doppler velocity, and Gaussian width) from all surge 
        and then, we averaged them for all the surges. Finally, we displayed both the correlations from these averaged parameters (see panel h: correlation between Doppler velocity and R$_{kh}$ and panel i: correlation between Gaussian width and R$_{kh}$). Overall,
        we find weak positive correlations between Doppler velocity and R$_{kh}$ (panel h) and a weak negative correlation between
        Gaussian width and R$_{kh}$ (panel i).} 
 \label{fig:kh_ratio}
\end{figure*}  

We have estimated R$_{kh}$ for the complete observation. And, R$_{kh}$ map along with Doppler velocity and
Gaussian sigma maps are shown in panel a, b, and c of Fig.~\ref{fig:kh_ratio}, respectively. First, we investigated 
the correlation among R$_{kh}$, Doppler velocity, and Gaussian sigma for S1. R$_{kh}$ is found to be low 
(i.e,~1.4) for the initial phase of S1 and it increases as time progresses. Blueshift in S1 decreases with the time and 
inverts into redshifts around the midtime for S1. This redehift further increases with time. Interestingly, R$_{kh}$ is positivley 
correlated with the Doppler velocity pattern (i.e., blueshifts decrease with time, and finally inverts 
into redshifts followed by the increase in Doppler velocity.)
Gaussian width is, 
however, negativley correlated with R$_{kh}$, i.e., R$_{kh}$ increases with decreasing Gaussian width. Panels (d) and (e) of Fig.~\ref{fig:kh_ratio} 
show the correlation between R$_{kh}$ and Doppler velocity for S1 and S2, respectively. From the figures it is evident that R$_{kh}$ 
and Doppler velocity are tightly $\&$ positively correlated, as Pearson's coefficents for S1 and S2 are 0.887 and 0.886, respectively. 
Correlation between R$_{kh}$ and Gaussian width for S1 and S2 are shown in panel (f) and (g) of Fig.~\ref{fig:kh_ratio}, respectively.
These two parameters are negatively correlated and Pearson's coefficent for S1 and S2 are -0.77 and -0.38, respectively.
 
For the final correlation plots, parameters (i.e., R$_{kh}$, Doppler velocity, and Gaussian width) were averaged over all 
the y-positions at each instant of time of all six surges. Hence, we got averaged array of each parameter that includes all
surges. Finally, panel (g) and panel (i) of Fig.~\ref{fig:kh_ratio} shows the correlation between R$_{kh}$ $\&$ Doppler velocity, and R$_{kh}$ and 
Gaussian width, respectively, averaged over all the surges. Correlation between R$_{kh}$ and Doppler velocity, averaged over all the 
surges, is positive and has the Pearson's coefficent value of 0.227. Correlation between R$_{kh}$ and Gaussian width, averaged over 
all the surges, is negative and has the Pearson's coefficent value of -0.18. 

\begin{figure*}
    \mbox{
 \hspace{-2.0cm}
    \includegraphics[trim = 4.0cm 3.0cm 4.5cm 0.0cm,scale=1.2]{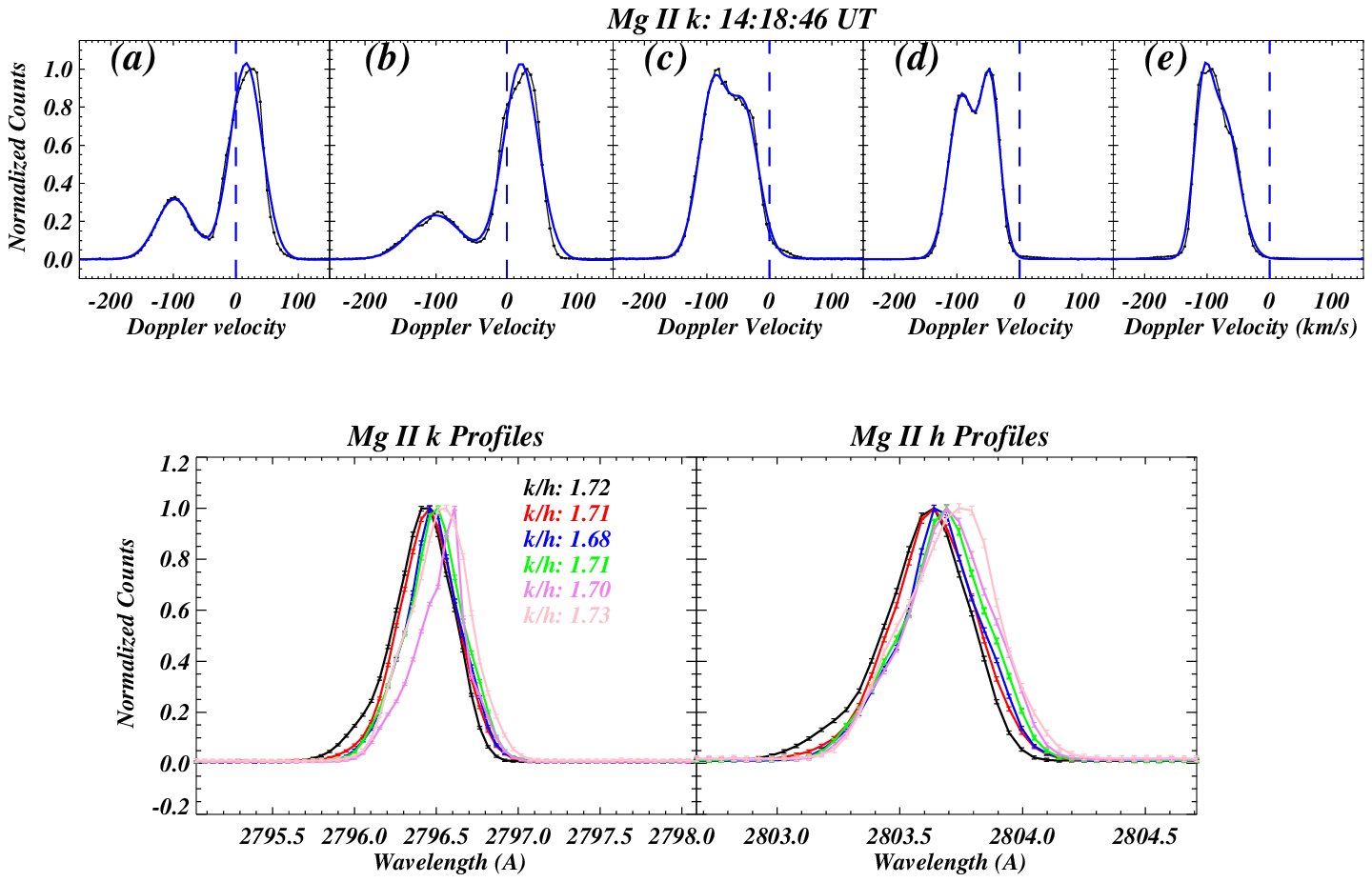}
}
  \caption{Top-row shows Mg~{\sc ii} k line taken from five different y-positions (i.e., same locations as displayed by horizontal 
slits by various colors in the panel A of figure~\ref{fig:hm_one}) during the evolution of S6 at T = 14:18:46~UT. We see a double peak profile for  Mg~{\sc ii} k line. 
One peak is constant while the other is shifting from red-shift at one edge of S6 (first-panel) to blue-shift at another edge of S6 (last-panel).  
This component is varying or non-constant component as defined already. 
In the bottom row, we have shown some Mg~{\sc ii} k (bottom-left panel) and Mg~{\sc ii} h (bottom-right panel) with different 
colors from the later phase of S6. In addition, we have mentioned also the R$_{kh}$ values for these spectral profiles. It is clearly visible 
that Mg~{\sc ii} k $\&$ h profiles are a single peak during the optically thick conditions.} 
 \label{fig:mg_evol}
\end{figure*}  

Mg~{\sc ii} k 2796.35~{\AA} line generally gets observed in the optically thick condition. Therefore, the complex behavior of Mg~{\sc ii} k 
line (i.e, double, triple, or multi-peak profile) do exist in the solar atmosphere. Although, in many circumstances Mg~{\sc ii} 
lines are single peak profiles, for instance, sunspot umbrae (\citealt{Tian2014}), solar flares (\citealt{Kerr2015}), and filaments 
(\citealt{Hara2014}). In addition, outside of the limb, Mg~{\sc ii} k lines exhibit single-peak profile (\citealt{Tei2020}). 
Therefore, in present scenario as spectra was taken far away from the limb (see IRIS slit in the left-panel of 
figure~\ref{fig:surge_spectra} {--} white dashed line), we believe that Mg~{\sc ii} lines should be single peaked profiles. 
Data obtained for Mg~{\sc ii} k 2796.35~{\AA} in S6 exhibit two peaks (see; top-panel of figure~\ref{fig:mg_evol}). It can be noted 
that the double peak profile exists only in the initial phase of surges, which becomes single peak profile in the
later phases of surges. We have already established the rotating motion for all the surges (see; figure~\ref{fig:hm_one} $\&$~\ref{fig:rot_s1_s5}) 
in TR. Rotating motion leads to the double peak in Si~{\sc iv} and O~{\sc iv} emission (i.e., optically thin lines). In principle, we 
should also see the rotating motion in chromospheric plasma,i.e., emission captured by Mg~{\sc ii} k $\&$ h lines. 
Mg~{\sc ii} k 2796.35~{\AA} profiles across S6 from time 14:18:46~UT are displayed in the top-panel of figure~\ref{fig:mg_evol}.
At the bottom-edge of S6, we observe peaks in the blue- and red-regimes of Mg~{\sc ii} k 2796.35~{\AA} line (panel a). Similar
Mg~{\sc ii} 2796.35~{\AA} profile exists in another location a little far away from the bottom-edge (panel b). However, in panel c, both peaks 
in Mg~{\sc ii} line have shifted to the blue-regime. Further, last two panels (i.e., panel d and e) also exhibit similar behaviour. 
The profiles of panel d and e are taken from near the upper edge of S6. Both the peaks of Mg~{\sc ii} line, 
at the upper edge of S6, lie in the blue-regime. Here also, the postion of one peak (in blue-regime) is constant (i.e., constant component) 
while other peak changes from redshifts at the lower edge of S6 to blueshifts at the upper edge of S6 (i.e., varying or
non-constant component). This behavior is similar to that of Si~{\sc iv} and O~{\sc iv} (see subsection~\ref{section:si_evol}) emissions. Thus, it can be 
concluded that double peak Mg~{\sc ii} profiles also aries due to simultaneous presence of rotational and translational motions in the 
surge plasma. 

Rotating motion of the surge plasma column is not a stable feature. It decays with time and vanishes after some time in each surge, 
e.g, after $\sim$15.0 minutes from the origin for S6. Thus we only see single peaks in Mg~{\sc ii} k $\&$ h profiles in the later 
phase of the surges (as shown for S6  in the bottom row of the figure~\ref{fig:mg_evol}). Please note that even for the single peak 
profiles, Mg~{\sc ii} k $\&$ h lines are still getting formed in optically thick-regime (i.e., R$_{kh}$ values are around $\sim$ 1.72 for 
the profiles displayed in bottom-panel of figure~\ref{fig:mg_evol}). 

\section{Discussion and Conclusions} 
      \label{sect:discussion}   

We utilized high-resolution SDO/AIA and IRIS observations to study six homologous surges that originated from AR 12114, located at 
the limb. These six homologous surges were originated within the time-span of $\sim$ 4 hours on 7$^{th}$ July 2014 from AR 12114. 
In the present study we started with the estimation of the kinematics of all surges using IRIS and SDO/AIA imaging observations. 
Variations in the lifetimes, up-flow and down-flow speeds, acceleration, and deceleration were found for these surges. Lifetime 
of surges varies from 23.0 to 50.0 minutes. Up-flow speed varies from 91.66 km s$^{-1}$ to 145.83 km s$^{-1}$. Acceleration varies 
from 152.97 m s$^{-2}$ to 477.27 m s$^{-2}$. Down-flow speed varies from 76.67 km s$^{-1}$
to 165.52 km s$^{-1}$. Deceleration varies from 127.76 ms$^{-2}$ to 667.08 m s$^{-2}$. Thus, it can be said that 
these homologous surges exhibit a large variation in kinematics. 

DEM of the surge's base shows the presence of two different temperature peaks, i.e., cool temperature distribution {--} log T/K = 
6.0 - 6.3, and hot temperature distribution {--} log T/K = 6.6 - 7.0. While, the spire (a region very far away from 
the base) of all surges emits only in the cool temperature regime (i.e., log T/K = 6.0 - 6.3 K). Thus, DEM analysis provides an important finding 
that there is existance of heated plasma at the origin site of the surges. 

The most striking feature of the present observation baseline is finding the emission of the surges in Si~{\sc iv}, O~{\sc iv},
and Mg~{\sc ii} k $\&$ h lines. Majority of the previous works have reported the surge emission in H$_{\alpha}$ or Ca~{\sc ii} 
8542~{\AA} (e.g., \citealt{Sch1995,Canfield1996,Kim2015,Huang2017,Yang2019}). Recently, \cite{Nob2017} have reported the emission 
of surge in Si~{\sc iv} only with the help of IRIS observations. However, the present observations provide a broader range of 
surge's emission, i.e., emission of surges in Mg~{\sc ii} k $\&$ h, Si~{\sc iv} and O~{\sc iv} lines. Here 
we report the surge's emission in two TR lines (i.e., Si~{\sc iv}, and O~{\sc iv}) and chromospheric 
Mg~{\sc ii} resonance lines (i.e., k $\&$ h). Thus, this is the  first report on the surge's emission in Mg~{\sc ii} resonance 
lines and O~{\sc iv}, and second work that report surge's emission in Si~{\sc iv} after \cite{Nob2017}. 

Shape of the spectral profiles is crucial as it provides important information about operating/on-going dynamics and physical 
process(es) within the in-situ plasma, e.g., bi-directional flows lead two distinct satellite lines during the explosive events 
(\cite{Dere1989}), enhanced wing emission in TR spectral lines above the networks that leads asymmetries 
in spectral profiles (\citealt{Peter2000}), several mechanisms may introduce the asymmetry in the coronal extreme ultraviolet 
lines originated within the active-regions (\citealt{Peter2010}), broadening of profiles due to hot explosion 
(\citealt{Peter2014}), and asymmetric profiles because of the prevalence of twist in the chromosphere/TR (\citealt{DePon2014}). 
\cite{Peter2000} and \cite{Peter2010} have demonstrated that double-Gaussian yields a reliable fitting 
for TR/coronal spectral profiles. 

IRIS provides the time-series spectral observations from the spire of surges under consideration. The v-t diagrams exhibit a 
systematic shift of Mg~{\sc ii} and Si~{\sc iv} lines from blue-shifts to red-shifts within the life-time of surges. 
The blueshifts means that the plasma is flowing upward, while redshift means plasma falling back towards 
the solar surface. The height-time diagrams (produced using imaging observations) also show that plasma is propagating in the upward 
direction in the first half of life-time while plasma is falling back towards the solar surface in the latter
half of life-time of each surge. We do see the up-flow and downflow episodes of surges from the height-time diagram produced 
using imaging observations from IRIS and AIA. The similar dynamics of surges can also be inferred from the 
v-t diagrams of Mg~{\sc ii} and Si~{\sc iv} lines.

Further, it is observed that the spectra of all lines for initial phase of each surge exhibits two peaks. Here, one of the component is 
constant (i.e., spectral line center does not change across the surges, for example, it is $\sim$ -100 km s$^{-1}$ 
in case of S6), while other component is varying (non-constant) across the surges (i.e., moves from the red-shifts to blue-shifts as 
we move from one edge of the surge to another edge). This is true for both, optically thin (i.e, Si~{\sc iv} and O~{\sc iv}) and optically thick (i.e., Mg~{\sc ii} k) lines. 
The constant component occurs due to the translational motion while the varying or non-constant is a result of rotational plasma.
This the varying component changes from redshifts to blueshifts as we move from one edge of the surge to another edge, which is a typical 
signature of the rotating plasma. Rotating plasma has also been reported in other jet-like structures of solar origin (e.g., \citealt[ and 
refernces cited therein]{Pike1998,Curdt2011,DePon2014,Young2014,Cheung2015b,Kayshap2018a}).

Magnetic reconnection process, that converts magnetic energy into thermal energy, may happen due to several activieits, e.g., flux-emergence, 
moving magnetic features, convective motions, etc. These processes are the fundamental base for majority of numerical 
simulations exhibiting the formation of jet-like structures within the solar atmosphere (e.g., \citealt{Yokoyama1995, Yokoyama1996, 
Nisizuka2008, Pariat2010, Srivastava2011, Kayshap2013b, Jelinek2015,Wyper2017}). More specifically, numerical simulations based on 
the magnetic reconnection between the twisted and pre-existing simpler magnetic field leads to the helical or
twisted field lines. Flow of plasma through these helical/twisted magnetic field lines appears as the rotating motion of the 
jet-plasma column (\citealt[and references cited therein]{Pariat2009, Pariat2016, Fang2014, Cheung2015}). And, energy produced in the 
magnetic reconnection process heats the plasma, that appears as the brightened/heated base for these jet-like structures in the 
observations (e.g., \citealt{Shibata1992,Yokoyama1995,Uddin2012,Kayshap2013a,Kayshap2013b,Kayshap2018b}). We have already established 
that the bases of the observed surges are heated, i.e., hot temperature component (Log T/K = 6.6-7.0) exists in the DEM of surge's base 
(figure~\ref{fig:dem_base_spire}). Thus, our analysis (e.g., heated plasma at the base of surges and rotating/helical motion) indicates 
the occurrence of the magnetic reconnection in the support of the formation of homologous surges as reported previously in various 
works (e.g., \citealt[and references cited therein]{Sch1995, Brooks2007,  Uddin2012, Kayshap2013a}). The amount of the released 
energy can vary from case to case, as the magnetic field and plasma conditions may vary spatially and temporally. Therefore, 
variations in the released energy during these homologous surges can be expected which leads to a large variation in the 
kinematic parameters (e.g., life-time, up-flow/down-flow, and velocity) of these surges.

Usually, double (or multi) peak exist in the Mg~{\sc ii} k 2796.35~{\AA} profiles for quiet-Sun and coronal holes (\citealt{Peter2014,Kayshap2018b}). 
However, Mg~{\sc ii} profiles are observed without the central reversal feature (i.e., single peak) in the flaring region, yet the line is 
optically thick (\citealt{Kerr2015}). Sunspot umbrae also show the single peak profiles despite the optically thick nature of Mg~{\sc ii} 
lines (\citealt{Morrill2001}). The current umbral models, assuming optically thick conditions, can not produce the single peak Mg~{\sc ii} k 2796.35~{\AA} profiles. 
Usually, decoupling between source function and Planck function in mid-chromosphere leads to the formation of central reversal or double 
peak profiles (\citealt{Leen2013}). However, for high-density plasma, coupling between line source and Planck function can continue even 
after the mid chromosphere. Thus, in such scenarios single profiles are observed (\citealt{Kerr2015}).

Generally outside the limb, Mg~{\sc ii} profiles exhibit a single peak (e.g., \citealt{Feldman1977, Vial1981, Hara2014}). Furthermore, 
spectral lines become narrower as we move away from the limb. However, in our case, we see both the double (initial phase) as well as a single 
peak (in the later phases of surges) profiles for surges. Here, double-peak profiles arise due to simultaneous presence of rotational 
and translational motion in the surges plasma, and not because of 
an optically thick atmosphere. In the later phase of surges (as rotation vanishes with time), for optically thick Mg~{\sc ii}, single peak 
Mg~{\sc ii} k $\&$ h profiles are observed. Optically thick conditions were revealed by R$_{kh}$ (i.e., R$_{kh}$ < 2.0). Please note that 
R$_{kh}$ has been measured for different plasma conditions,  e.g., R$_{kh}$ = 1.14-1.46 in various features (\citealt{Kohl1976,Lem1981,Lem1984}),
R$_{kh}$ = 1.20$\pm$0.010 in the non-flaring regions (\citealt{Kerr2015}), R$_{kh}$ = 1.07{--}1.19 in the 
region surrounding the flare emissions (before, during, and after the flare; \citealt{Kerr2015}), and R$_{kh}$ = 1.40{--}1.70 
in the filament material (\citealt{Hara2014}). In present work, 
averaged R$_{kh}$ varies from 1.40 to 2.20 in the surges (see; figure~\ref{fig:kh_ratio}). On an average, R$_{kh}$ value for surges seem to be higher as 
compared to that for other features. We have investigated the correlation of R$_{kh}$ with the Doppler velocity and 
width (see; Fig.~\ref{fig:kh_ratio}), also. R$_{kh}$ value increases with Doppler velocity pattern of surges, i.e., 
the blueshifts decrease with time, and inverts into the redshifts, followed by a increase in redshifts. R$_{kh}$ is positively correlated 
with the above described Doppler velocity pattern of surges, while R$_{kh}$ is negatively correlated with Gaussian width, i.e., R$_{kh}$ increases as Gaussian width 
decreases. It should be noted that Doppler velocity follows the above-described pattern and Gaussian width decreases as time progress for each surge. 
Hence, we conclude that R$_{kh}$ increases with the time as each surge progresses. It is to mention that \cite{Hara2014} have also reported 
an increase in the R$_{kh}$ following a filament eruption, for off-limb observations. 

Generally, a higher intensity line is an indicative of a density enhancement which can directly affect the shape of the line as mentioned by 
\cite{Kerr2015}. However, in case of optically thick lines, other mechanisms (e.g., radiation) can also populate the upper 
levels of resonance line.  \cite{Hara2014} reported that R$_{kh}$ would be closer to 4 if radiation is an important process in the particular in-situ plasma.
Hence, the absorption of radiation in the present scenario can be ruled out as we did not find such intensity ratios. Recent modeling, using 
bifrost (\citealt{Gudiken2011}) and the radiative transfer (\citealt{Per2015a}) codes, predicts that if there is a large temperature 
gradient ($\ge$1500 k) between the temperature minimum and formation region, Mg~{\sc ii} lines can undergo emission only (\citealt{Per2015b}).
In case of the flaring region, \cite{Kerr2015} reported the formation of single peak Mg~{\sc ii} profiles. However, it should be noted that 
\cite{Hara2014} has also found the single peak Mg~{\sc ii} profiles in cool filament above the limb. Further, the increment in R$_{kh}$ values 
occur after the plasma eruption (i.e., filament eruption/coronal mass ejection). Surges are also plasma ejection higher up in the atmosphere 
from the origin site. Thus, we believe that increase in R$_{kh}$ values with time is a result of plasma ejection, that in turn might be the 
result of some additional radiative excitation along with the collisional excitation as reported by \cite{Hara2014}. Finally, we mention that 
numerical modeling is needed to clarify this conclusion as also suggested by \cite{Hara2014}.  

In conclusion, after the rigorous investigation of the kinematic properties and thermal structure of six homologous surges, with the help of 
IRIS, and SDO observations, we have reported, for the first time, the emission of homologous surges in various spectral lines of the 
interface-region. It is found that all six homologous surges exhibit the rotating motion. Collectively, heated bases of the surge and rotating motion 
justify the occurrence of magnetic reconnection in support of the formation of surges. In the present work we have reported the time evolution
of R$_{kh}$, in case of surges, for the first time. 

\section*{acknowledgements}

IRIS is a small explorer mission developed and operated by LMSAL with mission operations executed at NASA Ames Research Center and 
major contributions to downlink communications funded by ESA and the Norwegian Space Center. SDO observations are courtesy of NASA/SDO 
and the aia, eve, and hmi science teams. CHIANTI is a collaborative project involving George Mason University (USA), University of Michigan 
(USA), University of Cambridge (UK) and NASA's Goddard Space Flight Center (USA).
\section*{Data Availability}
The data underlying this article are available at \url{https://iris.lmsal.com/data.html} (NASA/IRIS website) 
and at \url{https://iris.lmsal.com/search/} (LMSAL search website). Note that IRIS, as well as SDO/AIA data, 
are publically available with the observation id OBS 3864611254.


\bibliographystyle{mnras}
\bibliography{surge_pk_R4} 

\appendix \label{append:maps}
\section{Intensity, Doppler velocity and width maps: Mg~{\sc ii} k 2796.35~{\AA} and Si~{\sc iv} 1402.77~{\AA}}
\begin{figure*}	
    \mbox{
    \includegraphics[trim = 0.5cm 2.0cm 0.5cm 0.0cm,scale=0.72]{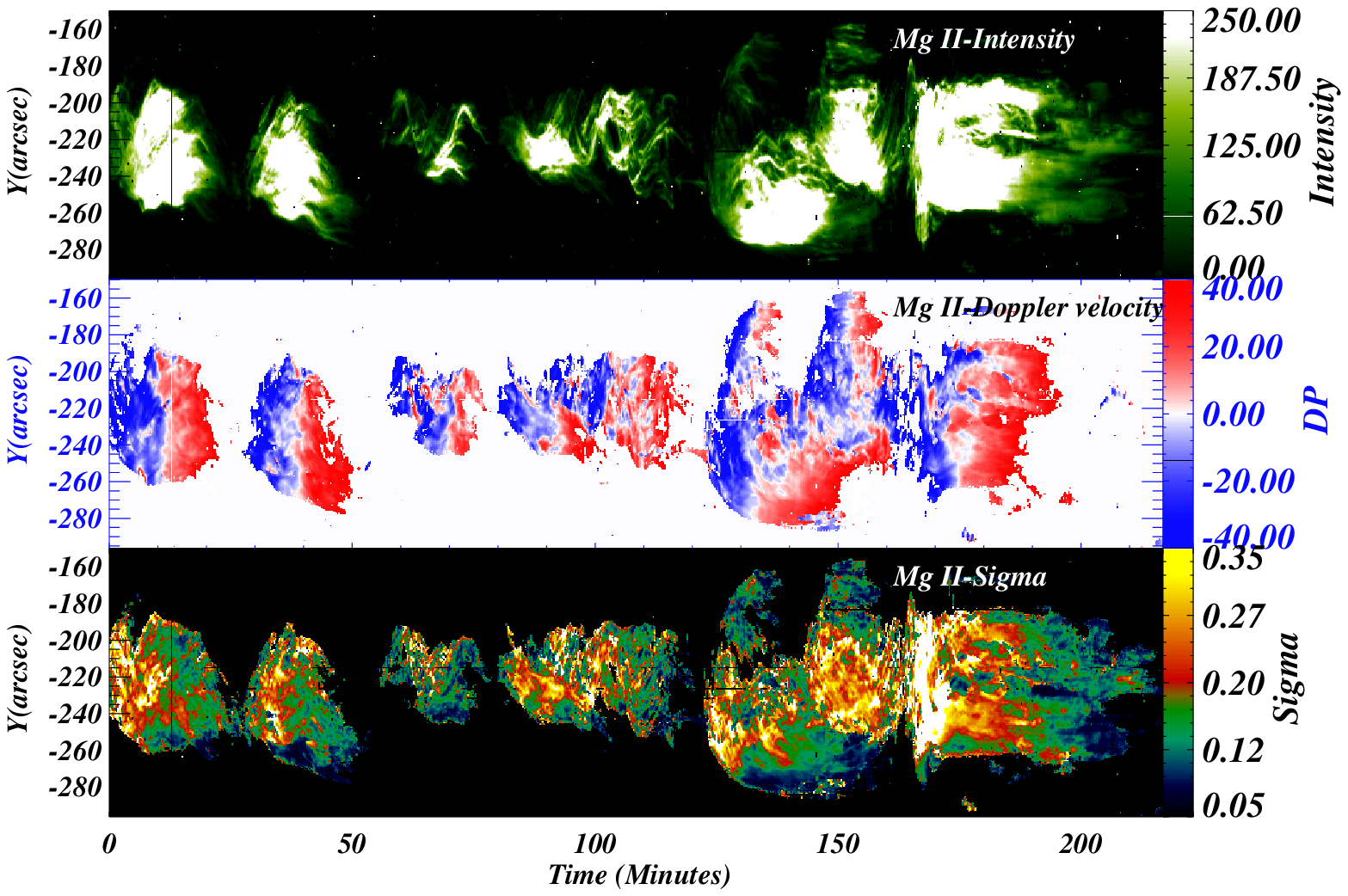}
    }
   \caption{Intensity, Doppler Velocity and Gaussian Sigma maps for the complete observation of Mg~{\sc ii} k 2796.35~{\AA}. 
	Displayed color bars in each panel show the range of corresponding parameter.}
\label{fig:mg_map_append}
\end{figure*}

\begin{figure*}	
    \mbox{
     \includegraphics[trim = 4.5cm 0.0cm 4.5cm 0.0cm,scale=0.72]{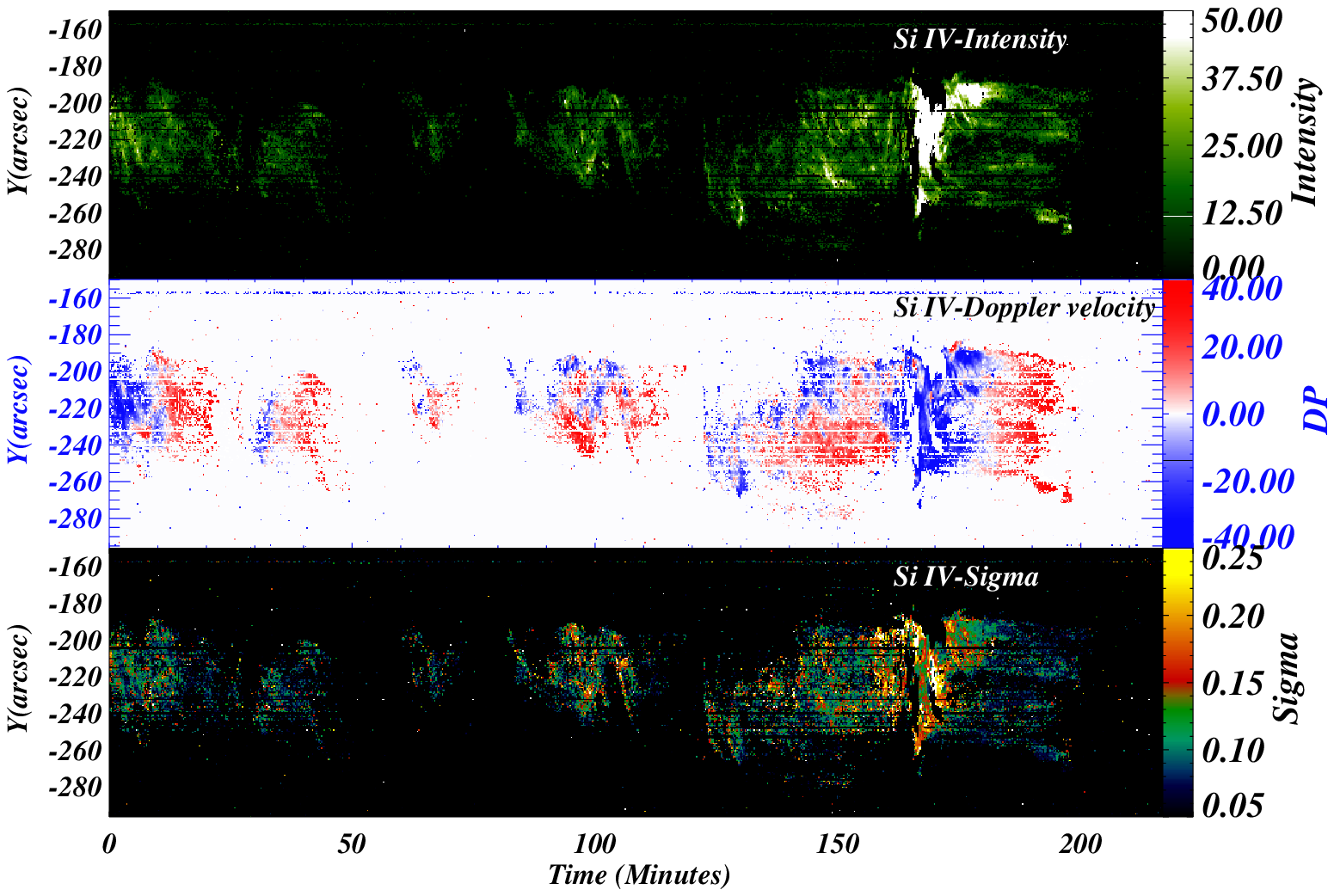}
     }
   \caption{same as figure~\ref{fig:mg_map_append} but for Si~{\sc iv} 1402.77~{\aa}.}
\label{fig:siv_map_append}
\end{figure*}

Intensity, Doppler velocity, and Gaussian width maps for Mg~{\sc ii} k 2796.35~{\AA} (fig.~\ref{fig:mg_map_append}) and 
Si~{\sc iv} 1402.77~{\AA} (fig.~\ref{fig:siv_map_append}) are shown in this appendix. Doppler velocity and Gaussian sigma 
maps for Mg~{\sc ii} k 2796.35~{\AA} and Si~{\sc iv} 1402.77~{\AA} were produced using the complete observations. Obtained 
spectra (for both Mg~{\sc ii} k 2796.35~{\AA} and Si~{\sc iv} 1402.77~{\AA}) were first fitted with single Gaussian function. 
Next, the Intensity, Doppler velocity, and Gaussian sigma of the complete observations were deduced using the single Gaussian fit. 
Single Gaussian fit was also used to produced these spectroscopic parameter maps for both lines. Doppler velocity maps show 
that almost the first half of each surge is dominated by blueshifts, while the later half is redshifted,  i.e., firstly the 
plasma is flowing up while after attaining the maximum height the plasma falls back to the solar surface. Such plasma dynamics 
(i.e., plasma up flow and downfall with time) is true for both Mg~{\sc ii} k and Si~{\sc iv} lines (see; middle-panels in 
fig.~\ref{fig:mg_map_append} and ~\ref{fig:siv_map_append}). Further, Gaussian width is higher in the initial phase of surges, 
it decrease with increasing time for both the lines Mg~{\sc ii} k and Si~{\sc iv} lines (see; lower-panels in 
fig.~\ref{fig:mg_map_append} and ~\ref{fig:siv_map_append})line. Thus, it can be said that the dynamics of surges is similar in 
both chromosphere (i.e., mg~{\sc ii} k 2796.35~{\AA}) and TR (i.e., Si~{\sc iv} 1402.77~{\AA}).


\section{Constant and varying (non-constant) components of Si~{\sc iv}: from first surge (S1) to fifth surge (S5)}
In case of most prominent surge (i.e., S6), we have already presented the dynamics of constant and varying or
non-constant components in both the lines (see; section~\ref{section:si_evol} and ~\ref{sect:mg_opacity}). We have also investigated the dynamics of constant
and varying or non-constant components for the rest of the surges, i.e., from S1 to S5. Behavior of both components were investigated by 
taking average of spectra over time-span of the surge emission. Averaged spectra were fitted using a double Gaussian function. Doppler 
velocity of both the components at each y-position was estimated on the basis of double Gaussian fitting. We utilize the same 
procedure to deduce the Doppler velocity of constant and varying components for all other surges. 

\label{append:evol_profile}
\begin{figure*}
    \mbox{
    \includegraphics[trim = 4.30cm 0.0cm 0.0cm 0.0cm,scale=0.50]{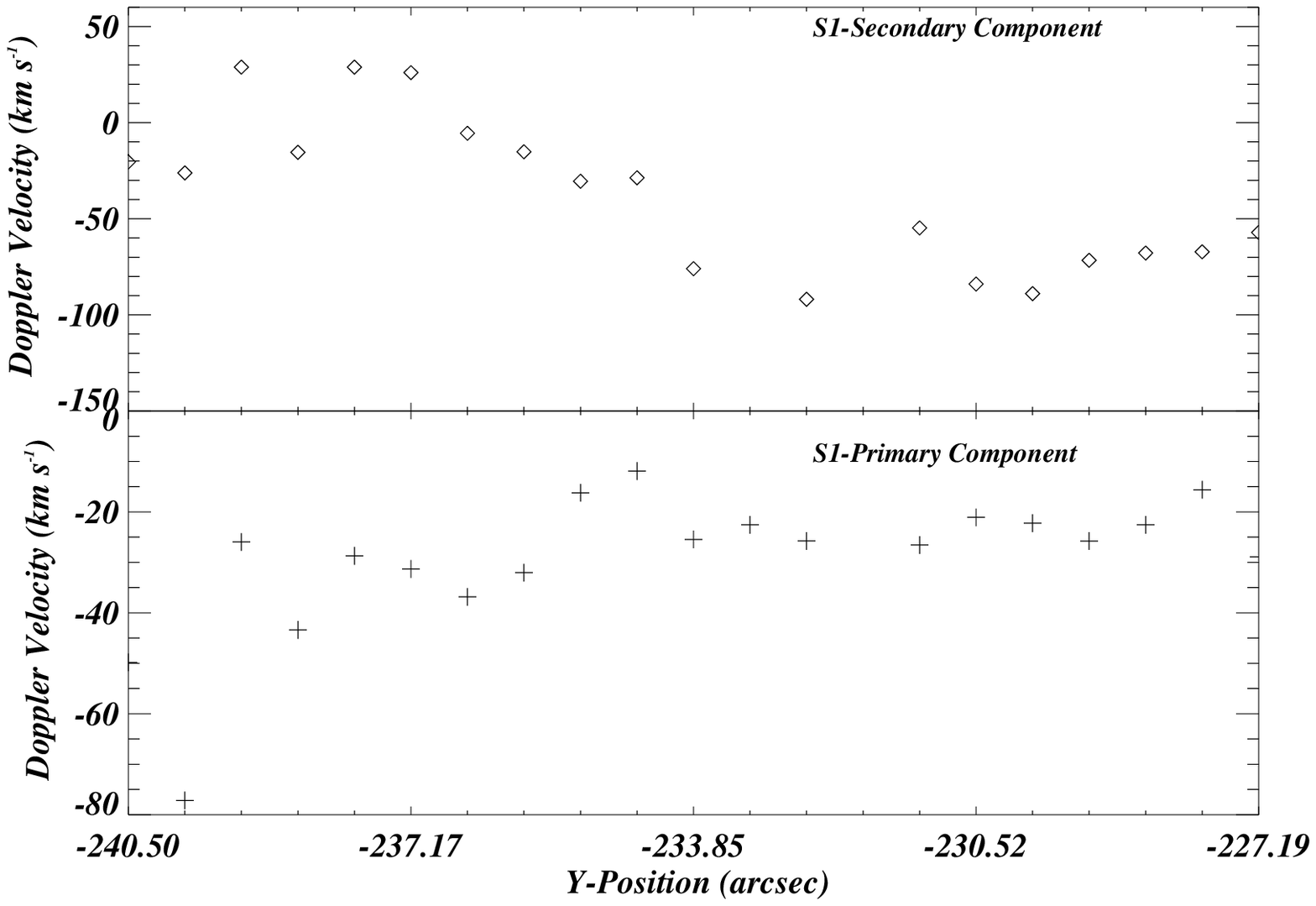}
    \includegraphics[trim = 2.30cm 0.0cm 0.5cm 0.0cm,scale=0.50]{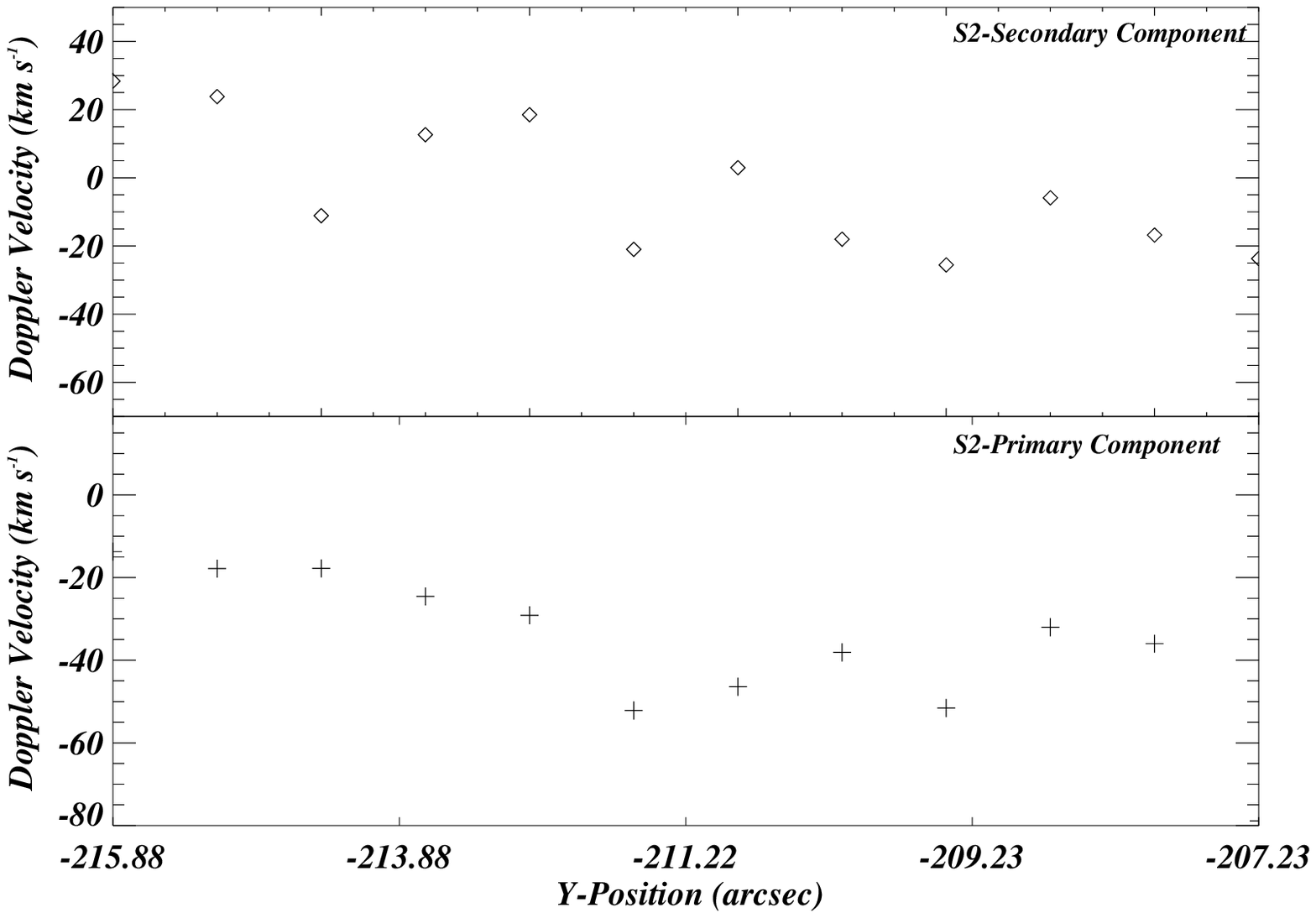}
}

    \mbox{
    \includegraphics[trim = 4.30cm 0.0cm 0.0cm 0.8cm,scale=0.50]{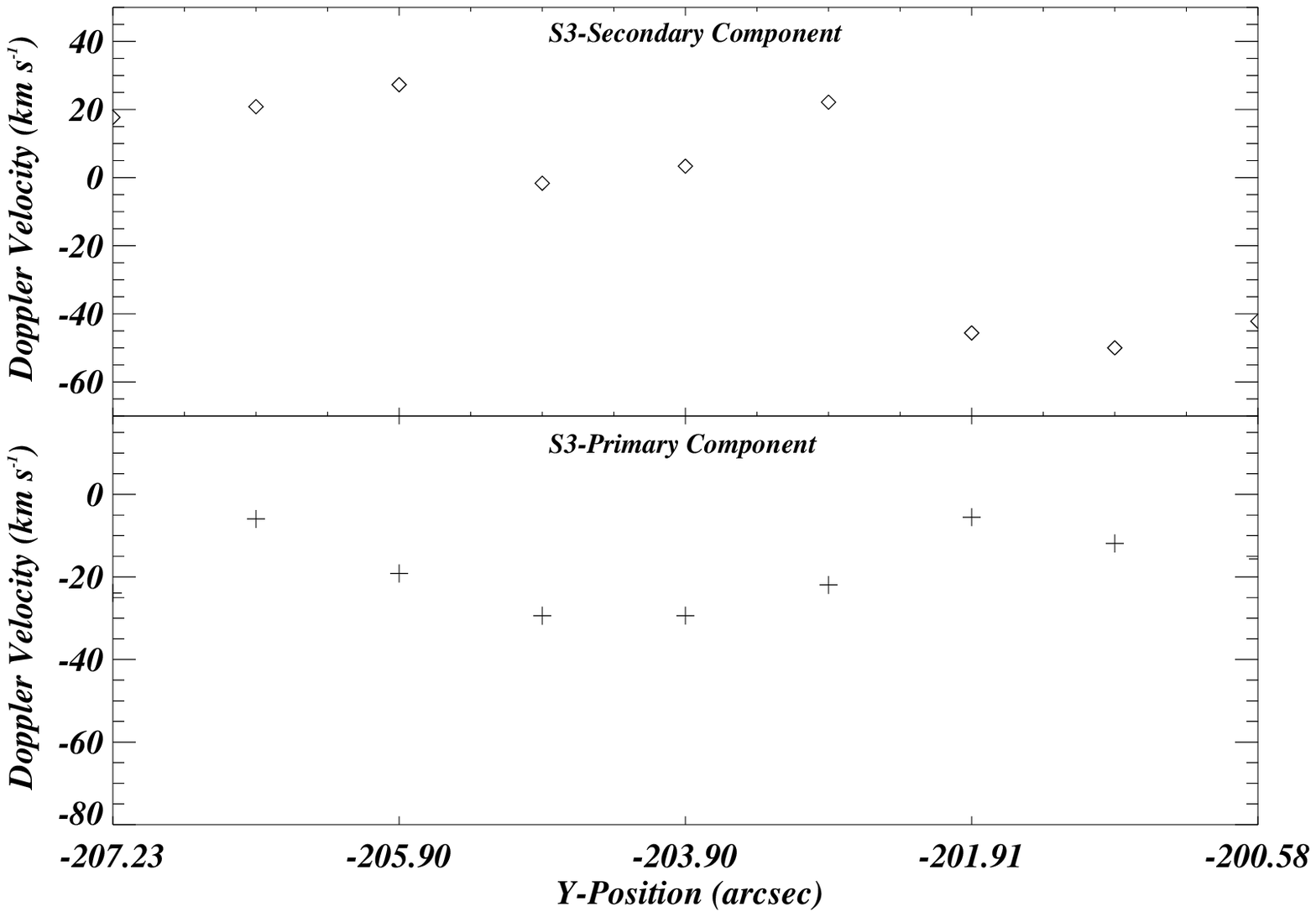}
    \includegraphics[trim = 2.30cm 0.0cm 0.5cm 0.8cm,scale=0.50]{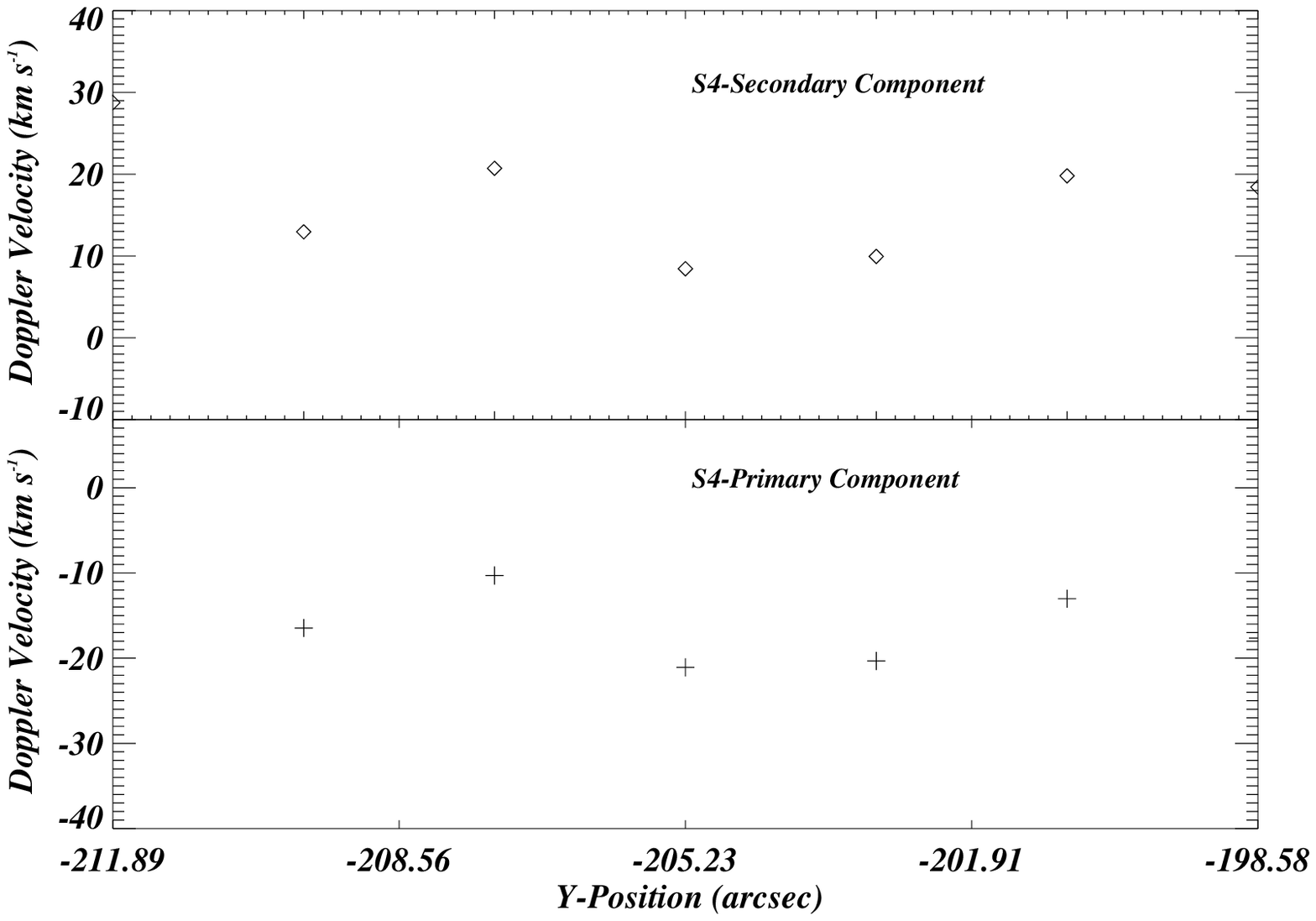}
}

 \mbox{
  
    \includegraphics[trim = -4.0cm 0.0cm 0.0cm 0.8cm,scale=0.50]{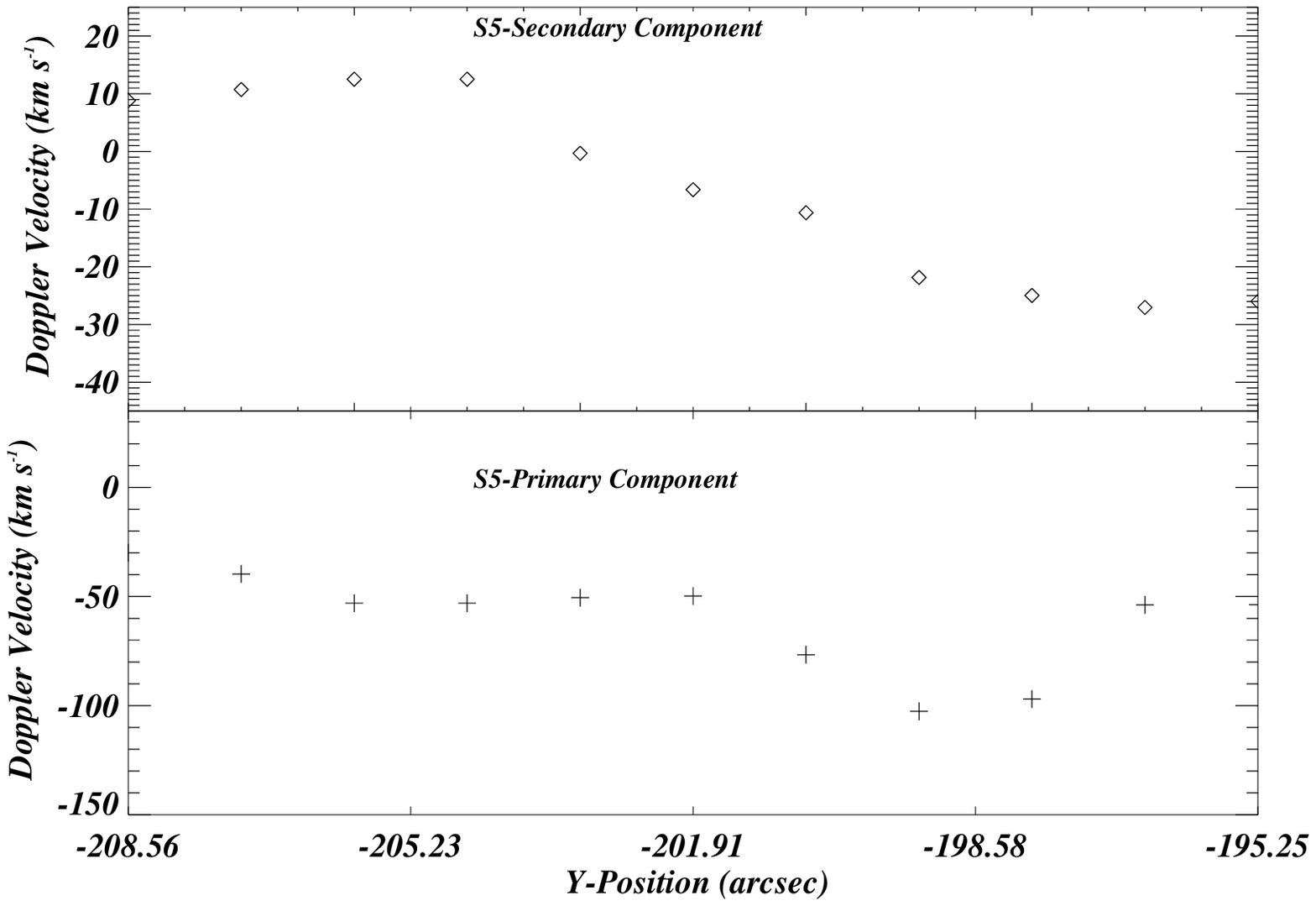}
}

   \caption{Here we show the behavior of primary and secondary components of Si~{\sc iv} line for S1 (top-left panel), S2 (top-right panel), 
	S3 (middle-left panel), S4 (middle-right panel), and S5(bottom panel). The primary component shows constant values 
	across all y-position in all surges, i.e., it is a constant component. While the secondary component shifts from 
	red-shift to blueshift in all surges, i.e., it is a varying component.}
 \label{fig:rot_s1_s5}
\end{figure*}  
At last, we have shown the Doppler velocity variations of constant and varying components for S1 (top-left panel), S2 (top-right panel), S3 (middle-left,
panel), S4 (middle-right panel), and S5 (bottom-panel) in figure~\ref{fig:rot_s1_s5}. In case of S1 (top-left panel), we do see that 
varying component is changing from redshift of $\sim$ 50 km s$^{-1}$ to blueshift of -100.0 km s$^{-1}$. 
While the constant component lies at $\sim$ -40.0 km s$^{-1}$ in all y-positions across the S1. Please note 
that similar type of results were found for other surges too (see other panels in fig.~\ref{fig:rot_s1_s5}). Thus, it can be concluded that 
for all the surges, one component shifts from the redshift to blueshift (i.e., varying component), while other component has constant 
velocity (i.e., non-constant).

\bsp	
\label{lastpage}
\end{document}